\documentclass[journal]{IEEEtran}
\usepackage{graphicx,cite,amssymb,amsmath,psfrag,subfigure}
\usepackage{graphicx}
\usepackage{subfigure}
\usepackage{mathrsfs}
\usepackage{color}
\usepackage{tabulary}
\usepackage{multirow}
\usepackage{cases}
\usepackage{stfloats}
\usepackage{url}
\usepackage[T1]{fontenc}

\renewcommand{\IEEEQED}{\IEEEQEDopen}

\newtheorem{theorem}{Theorem}

\newtheorem{proposition}{Proposition}

\newtheorem{remark}{Remark}
\newtheorem{algorithm}{Algorithm}

\DeclareMathOperator{\E}{\mathbb{E}}

\newcommand{\EX}[1]{\E\left\{{#1}\right\}}

\newcommand{\CG}[2]{\mathcal{CN}\left({#1},{#2}\right)}

\newcommand{\B}[1]{{\mathbf{#1}}}

\newcommand{\Pu}{\rho_{\mathrm{u}}}
\newcommand{\Pp}{\rho_{\mathrm{p}}}

\newcommand{\Pd}{\rho_{\mathrm{d}}}

\newcommand{\ul}{\mathrm{u}}
\newcommand{\p}{\mathrm{p}}
\newcommand{\dl}{\mathrm{d}}
\newcommand{\cf}{\mathrm{cf}}
\newcommand{\sce}{\mathrm{sc}}
\newcommand{\tauscul}{\tau^\mathrm{sc}_{\mathrm{u}}}
\newcommand{\tauscdl}{\tau^\mathrm{sc}_{\mathrm{d}}}
\newcommand{\taucf}{\tau^\mathrm{cf}}
\newcommand{\tauc}{\tau_\mathrm{c}}

\newcounter{eqncnt}

\newcounter{eqnback}

\begin{document}
\title{
    Cell-Free Massive MIMO versus Small Cells}
\author{{
        Hien Quoc Ngo, Alexei Ashikhmin, Hong Yang,  Erik G. Larsson, and Thomas L.
        Marzetta}
\thanks{Manuscript received August 03, 2015; revised February 22, 2016, August 25, 2016, and December 16, 2016; accepted
January 05, 2017. The associate editor coordinating the review of
this paper and approving it for publication was Dr.~Mai Vu.
The work of H.~Q.\ Ngo and E.~G.\ Larsson was
    supported in part by the Swedish Research Council (VR) and
    ELLIIT. Portions of this work were performed while H. Q. Ngo was with Bell Labs in 2014. Part of this work was presented at the 16th IEEE International
Workshop on Signal Processing Advances in Wireless Communications
(SPAWC) \cite{NAYLM:15:SPAWC}.}
\thanks{
        H.~Q.\ Ngo and E.~G. Larsson are with the Department of Electrical
    Engineering (ISY), Link\"{o}ping University, 581 83 Link\"{o}ping,
    Sweden
        (Email: hien.ngo@liu.se; erik.g.larsson@liu.se). H.~Q.\ Ngo is also with the School of Electronics, Electrical Engineering and Computer Science, Queen's University Belfast, Belfast BT3 9DT, U.K.
}

\thanks{A.\ Ashikhmin, H.\ Yang and T.~L.\ Marzetta are with Nokia Bell Laboratories, 
Murray Hill, NJ 07974 USA (Email: alexei.ashikhmin@nokia-bell-labs.com;
h.yang@nokia-bell-labs.com; tom.marzetta@nokia-bell-labs.com).}

\thanks{Digital Object Identifier xxx/xxx}
}

\markboth{IEEE Transactions on Wireless Communications, Vol. XX, No. X, XXX
2017}{Ngo \textit{\MakeLowercase{et al.}}: Cell-Free Massive MIMO versus Small Cells}

\maketitle

\begin{abstract}
A Cell-Free Massive MIMO (multiple-input multiple-output) system
comprises a very large number of distributed access points (APs)
which simultaneously serve a much smaller number of users over the
same time/frequency resources based on directly measured channel
characteristics. The APs and users have only one antenna each. The
APs acquire channel state information through time-division duplex
operation and the reception of uplink pilot signals transmitted by
the users. The APs perform multiplexing/de-multiplexing through
conjugate beamforming on the downlink and matched filtering on the
uplink. Closed-form expressions for individual user uplink and
downlink throughputs lead to  max-min power control algorithms.
Max-min power control ensures uniformly good service throughout
the area of coverage. A pilot assignment algorithm helps to
mitigate the effects of pilot contamination, but power control is
far more important in that regard.

Cell-Free Massive MIMO has considerably improved performance with
respect to a conventional small-cell scheme, whereby each user is
served by a dedicated AP, in terms of both 95\%-likely per-user
throughput and immunity to shadow fading spatial correlation.
Under uncorrelated shadow fading conditions, the cell-free scheme
provides nearly 5-fold improvement in 95\%-likely per-user
throughput over the small-cell scheme, and 10-fold improvement
when shadow fading is correlated.
\end{abstract}

\begin{IEEEkeywords}
Cell-Free Massive MIMO system,  conjugate
beamforming,  Massive MIMO, network MIMO, small cell.
\end{IEEEkeywords}

\section{Introduction} \label{Sec:Introduction}

\IEEEPARstart{M}{assive} multiple-input multiple-output (MIMO), where a base
station with many antennas simultaneously serves many users in the
same time-frequency resource, is a promising 5G wireless access
technology that can provide high throughput, reliability, and
energy efficiency with simple signal processing
\cite{Mar:10:WCOM,YM:14:VTC}. Massive antenna arrays at the base
stations can be deployed in collocated or distributed setups.
Collocated Massive MIMO architectures, where all service antennas
are located in a compact area, have the advantage of low backhaul
requirements. In contrast, in distributed Massive MIMO systems,
the service antennas are spread out over a large area.  Owing to
their ability to more efficiently exploit diversity against the
shadow fading, distributed systems can potentially offer much
higher probability of coverage than collocated Massive MIMO
\cite{ZZXWY:03:MCOM}, at the cost of increased backhaul
requirements.

In this work, we consider a distributed Massive MIMO
system where a large number of service antennas, called access
points (APs), serve a much smaller number of autonomous users
distributed over a wide area \cite{NAYLM:15:SPAWC}. All
APs cooperate phase-coherently via a backhaul network, and serve
all users in the same time-frequency resource via time-division
duplex (TDD) operation. There are no cells or cell boundaries.
Therefore, we call this system ``Cell-Free Massive MIMO''. Since
Cell-Free Massive MIMO  combines the distributed MIMO and Massive
MIMO concepts, it is expected to reap all benefits from these two
systems. In addition, since the users now are close to the APs,
Cell-Free Massive MIMO can offer a high coverage probability.
Conjugate beamforming/matched filtering techniques, also known as
maximum-ratio processing, are used both on uplink and downlink.
These techniques are computationally simple and can be implemented
in a distributed manner, that is, with most processing done
locally at the APs.\footnote{Other linear processing
techniques (e.g. zero-forcing) may improve the system performance,
but they require more backhaul than maximum-ratio processing does.
The tradeoff between the implementation complexity and the system
performance for these techniques is of interest and needs to be
studied in future work.}

In Cell-Free Massive MIMO, there is a central processing unit
(CPU), but the information exchange between the APs and this CPU
is limited to the payload data, and power control coefficients
that change slowly. There is no sharing of instantaneous channel
state information (CSI) among the APs or the central unit. All
channels are estimated at the APs through uplink pilots. The
so-obtained channel estimates are used to precode the transmitted
data in the downlink and to perform data detection in the uplink.
Throughout we emphasize per-user throughput rather than
sum-throughput. To that end we employ max-min power control.

In principle, Cell-Free Massive MIMO is an incarnation of general
ideas known as ``virtual MIMO'', ``network MIMO'', ``distributed
MIMO'', ``(coherent) cooperative multipoint joint processing'' (CoMP)
and ``distributed antenna systems'' (DAS).  The objective is to use
advanced backhaul to achieve coherent processing across geographically
distributed base station antennas, in order to provide uniformly good
service for all users in the network.  The outstanding aspect of
Cell-Free Massive MIMO is its operating regime: many single-antenna
access points simultaneously serve a much smaller number of users,
using computationally simple (conjugate beamforming) signal
processing. This facilitates the exploitation of phenomena such as
favorable propagation and channel hardening -- which are also key
characteristics of cellular Massive MIMO \cite{MLYN:16:Book}.  In turn, this
enables the use of computationally efficient and globally optimal
algorithms for power control, and simple schemes for pilot assignment
(as shown later in this paper).  In summary, Cell-Free Massive MIMO is
a useful and scalable implementation of the network MIMO and DAS
concepts -- much in the same way as cellular Massive MIMO is a useful
and scalable form of the original multiuser MIMO concept (see, e.g.,
\cite[Chap.~1]{MLYN:16:Book} for an extended discussion of the latter).

\textbf{Related work:}  

Many papers have studied network MIMO
\cite{KFV:06:WCM,IDMGFBMTJ:11:MCOM,HJWSG:13:JSAC} and DAS
\cite{CA:07:TWC,CG:10:WCM,HPWZ:13:COMM}, and indicated that network
MIMO and DAS may offer higher rates than colocated MIMO. However,
these works did not consider the case of very large numbers of service
antennas. Related works which use a similar system model as in our
paper are
\cite{LD:14:WCOM,SJWZGW:13:GLOBECOM,MZMR:13:JSAC,HWA:14:JSTSP,JCS:14:JSTSP,YBP:15:GLOBECOM,WD:15:WCOM}. In
these works, DAS with the use of many antennas, called large-scale DAS
or distributed massive MIMO, was exploited.  However, in all those
papers, perfect CSI was assumed at both the APs and the users, and in
addition, the analysis in \cite{WD:15:WCOM} was asymptotic in the
number of antennas and the number of users. A realistic analysis must
account for imperfect CSI, which is an inevitable consequence of the
finite channel coherence in a mobile system and which typically limits
the performance of any wireless system severely \cite{LHA:13:IT}.
Large-scale DAS with imperfect CSI was considered in
\cite{YGC:14:JSTSP,YM:13:ACCCC,TH:13:ACSSC,NAMY:15:ACSSC} for the
special case of orthogonal pilots or the reuse of orthogonal pilots, and in \cite{HTC:IT:12} assuming
frequency-division duplex (FDD) operation. In addition, in
\cite{YGC:14:JSTSP}, the authors exploited the low-rank structure of
users' channel covariance matrices, and examined the performance of
uplink transmission with matched-filtering detection, under the
assumption that all users use the same pilot sequence. By contrast, in
the current paper, we assume TDD operation, hence rely on reciprocity
to acquire CSI, and we assume the use of arbitrary pilot sequences in
the network -- resulting in pilot contamination, which was not studied
in previous work. We derive rigorous capacity lower bounds valid for
any finite number of APs and users, and give algorithms for optimal
power control (to global optimality) and pilot assignment.  

The papers cited above compare the performance between distributed
and collocated Massive MIMO systems. An alternative to
(distributed) MIMO systems is to deploy small cells, consisting of
APs that do not cooperate. Small-cell systems are
considerably simpler than Cell-Free Massive MIMO, since only data
and power control coefficients are exchanged between the CPU and
the APs. It is expected that Cell-Free Massive MIMO systems
perform better than small-cell systems.  However it is not clear, quantitatively,
 how much Cell-Free Massive MIMO systems can gain compared to
small-cell systems. Most previous work compares collocated Massive
MIMO and small-cell systems \cite{LHY:14:GlobalSIP,BSK:16:JSAC}.
In \cite{LHY:14:GlobalSIP}, the authors show that, when the number
of cells is large, a small-cell system is more energy-efficient
than a collocated Massive MIMO system. By taking into account a
specific transceiver hardware impairment and power consumption
model, paper \cite{BSK:16:JSAC} shows that reducing the cell size
(or increasing the base station density) is the way to increase
the energy efficiency. However when the circuit power dominates
over the transmission power, this benefit saturates. Energy
efficiency comparisons between collocated massive MIMO and
small-cell systems are also studied in
\cite{DKA:13:TWC,HWE:15:CL}.
 There has however been little work that
compares distributed Massive MIMO and small-cell systems. A
comparison between small-cell and distributed Massive MIMO systems
is reported in \cite{LD:14:WCOM},
 assuming
perfect CSI at both the APs and the users. Yet, a comprehensive
performance comparison between small-cell and distributed Massive
MIMO systems that takes into account the effects of imperfect CSI,
pilot assignment, and power control is not available in the
existing literature.

\textbf{Specific contributions of the paper:}

\begin{itemize}
\item We consider a cell-free massive MIMO with 
conjugate beamforming on the downlink and matched filtering on the
uplink. We show that, as in the case of collocated systems, when
the number of APs goes to infinity, the effects of non-coherent interference,
small-scale fading, and noise disappear.

\item We derive rigorous closed-form capacity lower bounds for the
Cell-Free Massive MIMO downlink and  uplink with finite numbers of
APs and users. Our analysis takes into account the effects of
channel estimation errors, power control, and non-orthogonality of
pilot sequences.

\item We compare two pilot assignment schemes: \emph{random
  assignment} and \emph{greedy assignment}.

\item We devise max-min fairness power control algorithms that maximize the
  smallest of all user rates. Globally
  optimal solutions can be computed by solving a sequence of
  second-order cone programs (SOCPs) for the downlink, and a sequence
  of linear programs for the uplink.

\item We quantitatively compare the performance of Cell-Free
Massive MIMO to that of  small-cell systems, under uncorrelated
and correlated shadow fading models.

\end{itemize}

The rest of paper is organized as follows. In
Section~\ref{Sec:SysModel}, we describe the Cell-Free Massive MIMO
system model. In Section~\ref{sec rate}, we present the achievable
downlink and uplink rates. The pilot assignment and power control
schemes are developed in Section~\ref{sec PA-PC}. The small-cell
system is discussed in Section~\ref{sec small cell}. We provide
numerical results and discussions in Section~\ref{sec
numerical-result} and finally conclude the paper in
Section~\ref{sec:Conclusion}.

\textit{Notation:}  Boldface letters denote column vectors. The
superscripts $()^\ast$, $()^T$, and $()^H$ stand for the
conjugate, transpose, and conjugate-transpose, respectively. The
Euclidean norm and the expectation operators are denoted by
$\|\cdot\|$ and $\mathbb{E}\left\{\cdot\right\}$, respectively.
Finally,  ${z} \sim \CG{0}{\sigma^2}$ denotes a circularly
symmetric complex Gaussian random variable (RV) ${z}$ with zero
mean and variance $\sigma^2$, and  ${z} \sim
\mathcal{N}(0,\sigma^2)$ denotes a real-valued Gaussian RV.

\begin{figure}[t!]
        \begin{center}
        \begin{psfrags}
        \psfrag{AP1}[tc][Bl][0.9]{AP $1$}
        \psfrag{AP2}[tc][Bl][0.9]{AP $2$}
        \psfrag{APm}[tc][Bl][0.9]{AP $m$}
        \psfrag{APM}[tc][Bl][0.9]{AP $M$}
        \psfrag{CPU}[tc][Bl][0.9]{CPU}
        \psfrag{terminal 1}[tc][Bl][0.90]{user $1$}
        \psfrag{terminal k}[tc][Bl][0.9]{user $k$}
        \psfrag{terminal K}[tc][Bl][0.9]{user $K$}
        \psfrag{gmk}[Bl][Bl][0.9]{$g_{mk}$}
        \includegraphics[width=0.50\textwidth]{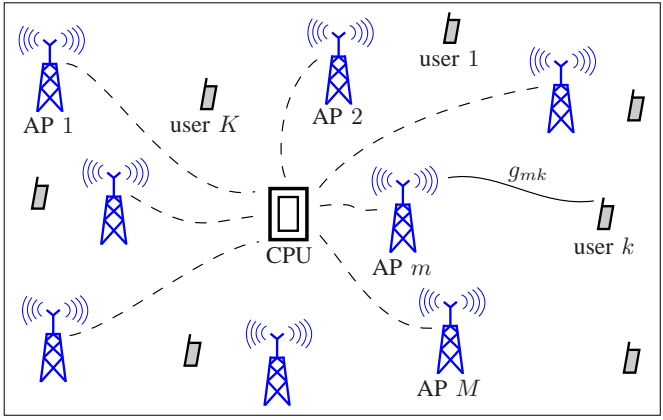}
        \end{psfrags}
\caption{Cell-Free Massive MIMO system.\label{fig:system}}
        \end{center}
\end{figure}

\section{Cell-Free Massive MIMO System Model} \label{Sec:SysModel}

We consider a Cell-Free Massive MIMO system with $M$ APs and $K$
users. All APs and users are equipped with a single antenna, and
they are randomly located in a large area. Furthermore, all APs
connect to a
  central processing unit via a backhaul network, see
  Figure~\ref{fig:system}. We assume that all $M$ APs simultaneously
serve all $K$ users in the same time-frequency resource. The
transmission from the APs to the users (downlink transmission) and
the transmission from the users to the APs (uplink transmission)
proceed by TDD operation. Each coherence interval is divided into
three phases: uplink training, downlink payload data transmission,
and uplink payload data transmission.  In the uplink training
phase, the users send pilot sequences to the APs and each AP
estimates the channel to all users. The so-obtained channel
estimates are used to precode the transmit signals in the
downlink, and to detect the signals transmitted from the users in
the uplink. In this work,  to avoid sharing of channel state
information between the APs, we consider conjugate beamforming in
the downlink and matched filtering in the uplink.

No pilots are transmitted in the
  downlink of Cell-Free Massive MIMO. The users do not need to
  estimate their effective channel gain, but instead rely on channel
  hardening, which makes this gain close to its expected value, a
  known deterministic constant. Our capacity bounds account for the
  error incurred when the users use the average effective channel
  gain instead of the actual effective gain. Channel hardening in
  Massive MIMO is discussed, for example, in \cite{Mar:10:WCOM}.

Notation is adopted and assumptions are made as follows:

\begin{itemize}
    \item The channel model
      incorporates the effects of small-scale fading and large-scale
      fading (that latter includes path loss and shadowing). The
      small-scale fading is assumed to be static during each coherence
      interval, and change independently from one coherence interval
      to the next. The large-scale fading changes much more slowly,
      and stays constant for several coherence intervals.
      Depending on the user mobility, the large-scale fading may stay
      constant for a duration of at least some $40$ small-scale fading
      coherence intervals \cite{Rapp:96:Book,AML:IT:14}.

    \item We assume that the channel is reciprocal, i.e., the channel
      gains on the uplink and on the downlink are the same.  This
      reciprocity assumption requires  TDD operation and perfect
      calibration of the hardware chains. The feasibility of the
      latter is demonstrated for example in \cite{Kaltenberger} for collocated Massive MIMO and it is conceivable that the problem can be similarly somehow for Cell-Free Massive MIMO.
      Investigating the effect of imperfect calibration is an important topic for future work.

    \item We let $g_{mk}$ denote the channel coefficient between the
      $k$th user and the $m$th AP. The channel $g_{mk}$ is modelled as
      follows:
\begin{align}\label{eq:gmk}
g_{mk} = \beta_{mk}^{1/2}h_{mk},
\end{align}
where $h_{mk}$ represents the small-scale fading, and $\beta_{mk}$
represents the large-scale fading. We assume that $h_{mk}$, $m=1,
\ldots, M$, $K=1, \ldots K$, are
  independent and identically distributed (i.i.d.) $\CG{0}{1}$ RVs.
The justification of the assumption of independent small-scale
fading is that the APs and the users are distributed over a wide
area, and hence, the set of scatterers is likely to be different
for each AP and each user.

\item We assume that all APs are
  connected via perfect backhaul that offers error-free and infinite
  capacity to the CPU. In practice, backhaul will be subject to
  significant practical constraints
  \cite{MF:10:WCOM,SSEPS:11:ITW}. Future work is needed to
  quantify the impact of backhaul constraints on
  performance.

\item In all scenarios, we let $q_k$ denote the   symbol
  associated with the $k$th user. These symbols are mutually
  independent, and independent of all noise and channel coefficients.
\end{itemize}

\subsection{Uplink Training} \label{sec CF CEst}

The Cell-Free Massive MIMO system employs a wide spectral
bandwidth, and the quantities $g_{mk}$ and $h_{mk}$ are dependent
on frequency; however $\beta_{mk}$ is constant with respect to
frequency. The propagation channels are assumed to be piece-wise
constant over a coherence time interval and a frequency coherence
interval. It is necessary to perform training within each such
time/frequency coherence block. We assume that $\beta_{mk}$ is
known, a priori, wherever required.

Let $\tauc$ be the length of the coherence interval (in samples),
which is equal to the product of the coherence time and the
coherence bandwidth, and let  $\taucf$ be the uplink training
duration (in samples) per coherence interval, where the
superscript $\cf$ stands for ``cell-free''. It is required that
$\taucf< \tauc$. During the training phase, all $K$ users
simultaneously send pilot sequences of length $\taucf$ samples to
the APs. Let $\sqrt{\taucf}\pmb{\varphi}_k \in
\mathbb{C}^{\taucf\times 1}$, where $\|\pmb{\varphi}_k\|^2=1$, be
the pilot sequence used by the $k$th user, $k=1, 2, \cdots, K$.
Then, the $\taucf\times 1$ received pilot vector at the $m$th AP
is given by
\begin{align}\label{eq:pilot1}
\B{y}_{\mathrm{p},m}  = \sqrt{\taucf
\Pp^{\mathrm{cf}}}\sum_{k=1}^K g_{mk} \pmb{\varphi}_k +
\B{w}_{\p,m},
\end{align}
where $\Pp^{\mathrm{cf}}$ is the normalized signal-to-noise ratio
(SNR) of each pilot symbol and $\B{w}_{\p,m}$ is a vector of
additive noise at the $m$th AP. The elements of $\B{w}_{\p,m}$ are
i.i.d. $\CG{0}{1}$ RVs.

Based on the received pilot signal $\B{y}_{\p,m}$, the $m$th AP
estimates the channel $g_{mk}, k=1, ..., K$. Denote by
$\check{y}_{\p,mk}$ the projection of $\B{y}_{\p,m}$ onto
$\pmb{\varphi}_k^H$:
\begin{align}\label{eq:tidley1}
&\check{y}_{\mathrm{p},mk} =  \pmb{\varphi}_k^H
\B{y}_{\mathrm{p},m}\nonumber\\ 
&= \sqrt{\taucf \Pp^{\mathrm{cf}}} g_{mk}  +
\sqrt{\taucf \Pp^{\mathrm{cf}}}\!\sum_{k'\neq k}^K g_{mk'}\!
\pmb{\varphi}_k^H \pmb{\varphi}_{k'} +
\pmb{\varphi}_k^H\B{w}_{\p,m}.
\end{align}
Although, for arbitrary pilot sequences, $\check{y}_{\p,mk}$ is
not a sufficient statistic for the estimation of $g_{mk}$, one can
still use this quantity to obtain suboptimal estimates. In the
special case when any two pilot sequences are either identical or
orthogonal, then $\check{y}_{\p,mk}$ is a sufficient statistic,
and estimates based on $\check{y}_{\p,mk}$ are optimal. The MMSE
estimate of $g_{mk}$ given $\check{y}_{\p,mk}$ is
\begin{align}\label{eq:MMSE est1}
\hat{g}_{mk} &=  \frac{\EX{\check{y}_{\p,mk}^\ast
g_{mk}}}{\EX{\left|\check{y}_{\p,mk} \right|^2}}
\check{y}_{\p,mk}= c_{mk}\check{y}_{\p,mk},
\end{align}
where $$c_{mk} \triangleq
\frac{\sqrt{\taucf\Pp^{\mathrm{cf}}}\beta_{mk}}{\taucf\Pp^{\mathrm{cf}}\sum_{k'=1}^K\beta_{mk'}\left|\pmb{\varphi}_k^H
\pmb{\varphi}_{k'}\right|^2+1}.$$

\begin{remark} \label{remark pilot}
If $\taucf \geq K$, then we can choose $\pmb{\varphi}_1,
\pmb{\varphi}_2, \cdots, \pmb{\varphi}_K$ so that they are
pairwisely orthogonal, and hence, the second term in
\eqref{eq:tidley1} disappears. Then the channel estimate
$\hat{g}_{mk}$ is independent of ${g}_{mk'}$, $k'\neq k$. However,
owing to the limited length of the coherence interval, in general,
$\taucf < K$, and mutually non-orthogonal pilot sequences must be
used throughout the network. The channel estimate $\hat{g}_{mk}$
is degraded by pilot signals transmitted from other users, owing
to the second term in \eqref{eq:tidley1}. This causes the
so-called pilot contamination effect.
\end{remark}

\begin{remark}
The channel estimation is performed in a decentralized fashion.
Each AP autonomously estimates the channels to the $K$ users. The
APs do not cooperate on the channel estimation, and no channel
estimates are interchanged among the APs.
\end{remark}

\subsection{Downlink Payload Data Transmission}

The APs treat the channel estimates as the true channels, and use
conjugate beamforming to transmit signals to the $K$ users. The
transmitted signal from the $m$th AP is given by
\begin{align}\label{eq:xm}
x_m = \sqrt{\Pd^{\mathrm{cf}}}\sum_{k=1}^K
\eta_{mk}^{1/2}\hat{g}_{mk}^\ast q_k,
\end{align}
where $q_k$, which satisfies $\EX{|q_k|^2}=1$, is the symbol
intended for the $k$th user, and $\eta_{mk}$, $m=1, \ldots, M$,
$k=1, \ldots K$, are power control coefficients chosen to satisfy
the following power constraint at each AP:
\begin{align}
\EX{|x_m|^2}\leq \Pd^{\mathrm{cf}}.
 \end{align}
With the channel model in \eqref{eq:gmk}, the power constraint
$\EX{|x_m|^2}\leq \Pd^{\mathrm{cf}}$ can be rewritten as:
\begin{align}\label{eq:pct}
\sum_{k=1}^K \eta_{mk}\gamma_{mk} \leq 1, \quad \text{for all
$m$},
\end{align}
where
\begin{align}\label{eq:gamma1}
\gamma_{mk}\triangleq \EX{\left|\hat{g}_{mk} \right|^2} =
\sqrt{\taucf\Pp^{\mathrm{cf}}}\beta_{mk}c_{mk}.
\end{align}

The received signal at the $k$th user is given by
\begin{align}\label{eq:rk1}
r_{\dl,k} 
&= 
\sum_{m=1}^M g_{mk}x_m +  w_{\dl,k}\nonumber\\
 &=
\sqrt{\Pd^{\mathrm{cf}}}\sum_{m=1}^M\sum_{k'=1}^K
\eta_{mk'}^{1/2}g_{mk}\hat{g}_{mk'}^\ast q_{k'} + w_{\dl,k},
\end{align}
where $w_{\dl,k}$ is additive $\CG{0}{1}$ noise at the $k$th user.
Then $q_k$ will be detected from $r_{\dl,k}$.

\subsection{Uplink Payload Data Transmission}

In the uplink, all $K$ users simultaneously send their data to the
APs. Before sending the data, the $k$th user weights its symbol
$q_k$, $\EX{|q_k|^2}=1$, by a power control coefficient
$\sqrt{\eta_k}$, $0\leq \eta_k \leq 1$. The received signal at the
$m$th AP is given by
\begin{align}\label{eq ULtrans 1}
y_{\ul,m} = \sqrt{\Pu^{\mathrm{cf}}} \sum_{k=1}^K
g_{mk}\sqrt{\eta_k}q_k + w_{\ul,m},
\end{align}
where $\Pu^{\mathrm{cf}}$ is the normalized uplink SNR and
$w_{\ul,m}$ is additive noise at the $m$th AP. We assume that
$w_{\ul,m}\sim \CG{0}{1}$.

To detect the symbol transmitted from
  the $k$th user, $q_k$, the $m$th AP multiplies the received signal
  $y_{\ul,m}$ with the conjugate of its (locally obtained) channel
  estimate $\hat{g}_{mk}$. Then the so-obtained quantity
  $\hat{g}_{mk}^\ast y_{\ul,m}$ is sent to the CPU via a backhaul
  network. The CPU sees
\begin{align}\label{eq ULtrans 21}
 r_{\ul,k} &= \sum_{m=1}^M\hat{g}_{mk}^\ast y_{\ul,m}\nonumber\\
 &=
 \sum_{k'=1}^K\sum_{m=1}^M \sqrt{\Pu^{\mathrm{cf}}\eta_{k'}}
 \hat{g}_{mk}^\ast g_{mk'} q_{k'} + \sum_{m=1}^M \hat{g}_{mk}^\ast
 w_{\ul,m}.
 \end{align}
Then, $q_k$ is detected from $ r_{\ul,k}$.

\section{Performance Analysis}\label{sec rate}

\subsection{Large-$M$ Analysis}\label{sec: largeM}

In this section, we provide some
  insights into the performance of Cell-Free Massive MIMO systems when
  $M$ is very large. The convergence analysis is done conditioned on a set of deterministic large-scale fading coefficients $\{\beta_{mk}\}$. We show that, as in the case of Collocated
  Massive MIMO, when $M\to\infty$, the channels between the users and
  the APs become orthogonal. Therefore, with conjugate beamforming
  respectively matched filtering, non-coherent interference,
  small-scale fading, and noise disappear. The only remaining
  impairment is pilot contamination, which consists of interference
  from users using same pilot sequences as the user of interest in the
  training phase.

On downlink, from \eqref{eq:rk1},
  the received signal at the $k$th user can be written as:
\begin{align}\label{eq:rk1M}
r_{\dl,k}
    &=
    \underbrace{\sqrt{\Pd^{\mathrm{cf}}}\sum_{m=1}^M \eta_{mk}^{1/2}g_{mk}\hat{g}_{mk}^\ast
    q_{k}}_{\text{DS}_k}\nonumber\\
    &+
    \underbrace{\sqrt{\Pd^{\mathrm{cf}}}\sum_{m=1}^M\sum_{k'\neq k}^K \eta_{mk'}^{1/2}g_{mk}\hat{g}_{mk'}^\ast
    q_{k'}}_{\text{MUI}_k}
    +
    w_{\dl,k},
\end{align}
where $\text{DS}_k$ and $\text{MUI}_k$ represent the desired
signal and multiuser interference, respectively.

By using the channel estimates in \eqref{eq:MMSE est1}, we have
\begin{align}\label{eq:rk2M}
&\sum_{m=1}^M \eta_{mk'}^{1/2}g_{mk}\hat{g}_{mk'}^\ast
    \nonumber\\
    &=
    \sum_{m=1}^M\! \eta_{mk'}^{1/2}c_{mk'}g_{mk}\!\!\left(\!\!\sqrt{\taucf \Pp^{\mathrm{cf}}}\sum_{k''=1}^K g_{mk''} \pmb{\varphi}_{k'}^H \pmb{\varphi}_{k''} + \tilde{w}_{\p,mk'}
    \!\right)^\ast\nonumber\\
    &=
    \sqrt{\taucf \Pp^{\mathrm{cf}}} \sum_{m=1}^M
    \eta_{mk'}^{1/2}c_{mk'}|g_{mk}|^2  \pmb{\varphi}_{k'}^T
    \pmb{\varphi}_{k}^\ast \nonumber\\
    &+ \sqrt{\taucf \Pp^{\mathrm{cf}}}\sum_{k''\neq k}^K \sum_{m=1}^M \eta_{mk'}^{1/2}c_{mk'}g_{mk} g_{mk''}^\ast \pmb{\varphi}_{k'}^T
    \pmb{\varphi}_{k''}^\ast\nonumber\\
    &+ \sum_{m=1}^M
    \eta_{mk'}^{1/2}c_{mk'}g_{mk}\tilde{w}_{\p,mk'}^\ast,
\end{align}
where $\tilde{w}_{\p,mk'} \triangleq
\pmb{\varphi}_{k'}^H\B{w}_{\p,m}$. Then by Tchebyshev's theorem
\cite{Cra:70:Book},\footnote{Tchebyshev's theorem: Let $X_1, X_2,
...X_n$ be independent RVs such that $\EX{X_i} = \mu_i$ and ${\tt
Var}\left\{X_i \right\}\leq c<\infty$, $\forall i$. Then
$$
\frac{1}{n}\left(X_1 + X_2 + ... + X_n\right)
-\frac{1}{n}\left(\mu_1 + \mu_2 + ... \mu_n\right) \mathop  \to
\limits^{P} 0.
$$
 }
 we have
\begin{align}\label{eq:rk3M}
    &\frac{1}{M}\! \sum_{m=1}^M \eta_{mk'}^{1/2}g_{mk}\hat{g}_{mk'}^\ast
    -\frac{1}{M}     \sqrt{\taucf \Pp^{\mathrm{cf}}} \sum_{m=1}^M
    \eta_{mk'}^{1/2}c_{mk'}\beta_{mk}  \pmb{\varphi}_{k'}^T
    \pmb{\varphi}_{k}^\ast  \nonumber\\
    &\mathop  \to \limits^{P}_{M \to\infty} ~
    0.
\end{align}
Using \eqref{eq:rk3M}, we obtain the following results:
\begin{align}\label{eq:rk4M}
    &\frac{1}{M}{\text{DS}}_k
    -\frac{1}{M}     \sqrt{\taucf \Pp^{\mathrm{cf}}\Pd^{\mathrm{cf}}} \sum_{m=1}^M
    \eta_{mk}^{1/2}c_{mk}\beta_{mk}
    q_{k} ~
    \mathop  \to \limits^{P}_{M \to\infty} ~
    0, \\
    &\frac{1}{M}{\text{MUI}}_k - \frac{1}{M} \sqrt{\taucf \Pp^{\mathrm{cf}}\Pd^{\mathrm{cf}}}\sum_{m=1}^M\sum_{k'\neq k}^K
    \eta_{mk'}^{1/2}c_{mk'}\beta_{mk}  \pmb{\varphi}_{k'}^T
    \pmb{\varphi}_{k}^\ast
    q_{k'}
     \nonumber\\
    &\mathop  \to \limits^{P}_{M \to\infty} ~
    0.
\end{align}

The above expressions show that when
  $M\to\infty$, the received signal includes only the desired signal
  plus interference originating from the pilot sequence
  non-orthogonality:
\begin{align}\label{eq:rk5M}
    &\frac{r_{\dl,k}}{M}
    -
    \frac{\sqrt{\taucf \Pp^{\mathrm{cf}}\Pd^{\mathrm{cf}}}}{M}\left(  \sum_{m=1}^M
    \eta_{mk}^{1/2}c_{mk}\beta_{mk}
    q_{k}\right.\nonumber\\
    &\left.
    +
     \sum_{m=1}^M\sum_{k'\neq k}^K
    \eta_{mk'}^{1/2}c_{mk'}\beta_{mk}  \pmb{\varphi}_{k'}^T
    \pmb{\varphi}_{k}^\ast
    q_{k'}\right) ~ \mathop  \to \limits^{P}_{M \to\infty} ~
    0.
\end{align}
If the pilot sequences are pairwisely orthogonal, i.e.,
$\pmb{\varphi}_{k'}^H \pmb{\varphi}_{k} =0$ for $k\neq k'$, then
the received signal becomes free of interference and noise:
\begin{align}\label{eq:rk6M}
    \frac{r_{\dl,k}}{M}
    -
    \frac{\sqrt{\taucf \Pp^{\mathrm{cf}}\Pd^{\mathrm{cf}}}}{M} \sum_{m=1}^M
    \eta_{mk}^{1/2}c_{mk}\beta_{mk}
    q_{k}
    ~ \mathop  \to \limits^{P}_{M \to\infty} ~
    0.
\end{align}
Similar results hold on the uplink.

\subsection{Achievable Rate for Finite $M$}

In this section, we derive closed-form expressions for the
downlink and uplink achievable rates, using the analysis technique
from \cite{HH:03:IT,YM:13:ACCCC}.

\subsubsection{Achievable Downlink Rate}\label{sec DLrate}

We assume that each user has knowledge of the channel statistics
but not of the channel realizations. The received signal
$r_{\dl,k}$ in \eqref{eq:rk1} can be written as
\begin{align}\label{eq:rate1}
    r_{\dl,k} =
  {\tt DS}_k\cdot q_{k}+
  {\tt BU}_k\cdot q_{k}
  +  \sum_{k'\neq k}^K
  {\tt UI}_{kk'}\cdot q_{k'} + w_{\dl,k},
\end{align}
where
\begin{align}\label{eq:rat2a}
    {\tt DS}_k &\triangleq  \sqrt{\Pd^{\mathrm{cf}}}\EX{\sum_{m=1}^M \eta_{mk}^{1/2}g_{mk}\hat{g}_{mk}^\ast},\\\label{eq:rat2b}
    {\tt BU}_k &\triangleq \sqrt{\Pd^{\mathrm{cf}}}\!\left(\sum_{m=1}^M
  \eta_{mk}^{1/2}g_{mk}\hat{g}_{mk}^\ast\!-\!
  \EX{\sum_{m=1}^M  \eta_{mk}^{1/2}g_{mk}\hat{g}_{mk}^\ast}\right),\\\label{eq:rat2c}
    {\tt UI}_{kk'} &\triangleq \sqrt{\Pd^{\mathrm{cf}}}\sum_{m=1}^M   \eta_{mk'}^{1/2}g_{mk}\hat{g}_{mk'}^\ast,
\end{align}
represent the strength of desired signal (DS), the beamforming
gain uncertainty (BU), and the interference caused by the $k'$th
user (UI), respectively.

We treat the sum of the second, third, and fourth terms in
\eqref{eq:rate1} as ``effective noise''. Since $q_k$ is
independent of ${\tt DS}_k$ and ${\tt BU}_k$, we have
$$
\EX{ {\tt DS}_k\cdot q_{k} \times \left({\tt BU}_k\cdot q_{k}
    \right)^\ast}
    =
    \EX{ {\tt DS}_k \times {\tt BU}_k^\ast} \EX{|q_k|^2}
    =
    0.$$ Thus, the first and the second terms of \eqref{eq:rate1} are
    uncorrelated. A similar calculation shows that the third and
    fourth terms of \eqref{eq:rate1} are uncorrelated with the first
    term of \eqref{eq:rate1}. Therefore, the effective noise and the
    desired signal are uncorrelated. By using the fact that
uncorrelated Gaussian noise represents the worst case, we obtain
the following achievable rate of the $k$th user for Cell-Free (cf)
operation:
\begin{align}\label{eq:rateexpr1}
R_{\dl,k}^{\cf}
 =
 \log_2
    \left(
    1 + \frac{\left|{\tt DS}_k \right|^2}{\EX{|{\tt BU}_k|^2} + \sum_{k'\neq k}^K  \EX{|{\tt UI}_{kk'}|^2} + 1}
    \right).
\end{align}

We next provide a new exact closed-form expression for the
achievable rate \eqref{eq:rateexpr1}, for a finite $M$.

\begin{theorem}\label{theorem rate}
An achievable downlink rate of the transmission from the APs to
the $k$th user in the Cell-Free Massive MIMO system with conjugate
beamforming, for any finite $M$ and $K$, is given by \eqref{eq:Theo_rateexpr1}, shown at the top of the next page.
\begin{IEEEproof}
See Appendix~\ref{sec app theo1}.
\end{IEEEproof}
\end{theorem}

\setcounter{eqnback}{\value{equation}} \setcounter{equation}{23}
\begin{figure*}[!t]
\begin{align}\label{eq:Theo_rateexpr1}
R_{\dl,k}^{\cf}
 =
 \log_2
    \left(
    1 + \frac{\Pd^{\mathrm{cf}} \left(\sum\limits_{m=1}^M \eta_{mk}^{1/2}\gamma_{mk} \right)^2 }{ \Pd^{\mathrm{cf}}\sum\limits_{k'\neq k}^K\left(\sum\limits_{m=1}^M \eta_{mk'}^{1/2}\gamma_{mk'}\frac{\beta_{mk}}{\beta_{mk'}} \right)^2| \pmb{\varphi}_{k'}^H\pmb{\varphi}_{k}|^2 + \Pd^{\mathrm{cf}}\sum\limits_{k'=1}^K\sum\limits_{m=1}^M \eta_{mk'}\gamma_{mk'}\beta_{mk} +1 }
    \right),
\end{align}
\hrulefill
\setcounter{equation}{26}\begin{align}\label{eq:Theo_uprateexpr1}
R_{\ul,k}^{\cf}
 =
 \log_2
    \left(
    1 + \frac{\Pu^{\mathrm{cf}} \eta_k\left(\sum\limits_{m=1}^M \gamma_{mk} \right)^2 }{ \Pu^{\mathrm{cf}}\!\sum\limits_{k'\neq k}^K \eta_{k'}\!\left(\sum\limits_{m=1}^M \gamma_{mk}\frac{\beta_{mk'}}{\beta_{mk}} \right)^2 |\pmb{\varphi}_{k}^H\pmb{\varphi}_{k'}|^2 + \Pu^{\mathrm{cf}}\!\sum\limits_{k'=1}^K\eta_{k'}\!\sum\limits_{m=1}^M \gamma_{mk}\beta_{mk'} +\sum\limits_{m=1}^M \gamma_{mk}}
    \!\right),
\end{align}
\hrulefill
\end{figure*}
\setcounter{eqncnt}{\value{equation}}
\setcounter{equation}{\value{eqnback}}

\begin{remark}
The main differences between the capacity bound expressions for
Cell-Free and collocated Massive MIMO systems \cite{YM:14:VTC}
are: i) in Cell-Free systems, in general $\beta_{mk}
\neq\beta_{m'k}$, for $m\neq m'$, whereas in collocated Massive
MIMO, $\beta_{mk} =\beta_{m'k}$; and ii) in Cell-Free systems, a
power constraint is applied at each AP individually, whereas in
collocated systems, a total power constraint is applied at each
base station. Consider the special case in which all APs are
collocated and the power constraint for each AP is replaced by a
total power constraint over all APs. In this case, we have
$\beta_{mk} =\beta_{m'k}\triangleq \beta_{k}$, $\gamma_{mk}
=\gamma_{m'k}\triangleq \gamma_{k}$, and the power control
coefficient is ${\eta}_{mk} =\eta_k/ (M\gamma_{mk})$. If,
furthermore, the $K$ pilot sequences are pairwisely orthogonal,
then, \eqref{eq:Theo_rateexpr1} becomes
\setcounter{equation}{24}\begin{align}\label{eq:rate spec}
R_{\dl,k}^{\cf}
 =
 \log_2
    \left(
    1 + \frac{M\Pd^{\mathrm{cf}} \gamma_k\eta_k }{ \Pd^{\mathrm{cf}}\beta_k\sum_{k'=1}^K\eta_{k'} +1 }
    \right),
\end{align}
which is identical to the rate expression for collocated Massive
MIMO systems in \cite{YM:14:VTC}.
\end{remark}

\begin{figure}[t!]
\centerline{\includegraphics[width=0.5\textwidth]{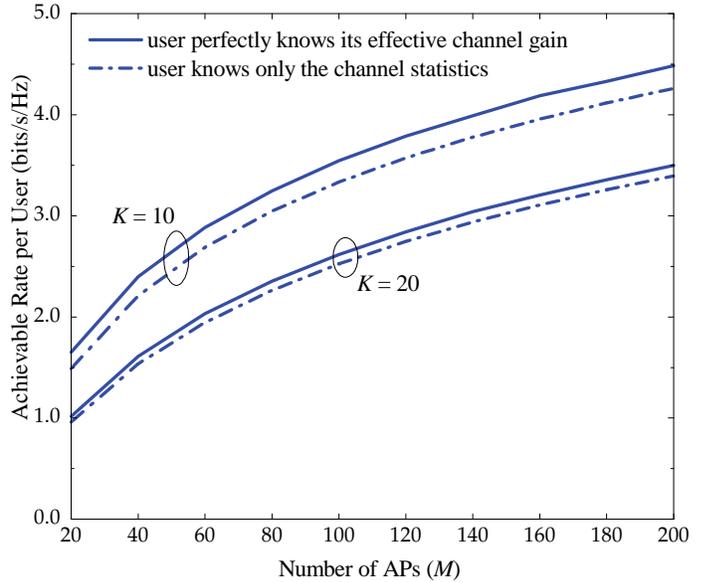}}
\caption{The achievable rate versus the number of APs for
different $K$. Here, $\Pd^\sce = 10$~dB, $\Pp^{\mathrm{cf}} =
0$~dB, $\taucf=K$, $\beta_{mk}=1$, $\eta_{mk} = 1/(K\gamma_{mk})$,
and pilot sequences are pairwisely orthogonal.\label{fig:harden}}
\end{figure}

\begin{remark}\label{remark_tightness}
The achievable rate \eqref{eq:Theo_rateexpr1} is obtained under
the assumption that the users only know the channel statistics.
However, this achievable rate is close to that in the case where
the users know the actual channel realizations. This is a
consequence of channel hardening, as discussed in
Section~\ref{Sec:SysModel}. To see this more quantitatively, we
compare the achievable rate \eqref{eq:Theo_rateexpr1} with the
following expression,
\begin{align}\label{eq aaa}
\tilde{R}_{\dl,k}^{\cf}
 =
 \EX{\log_2\!
    \left(\!
    1 + \frac{\Pd^{\mathrm{cf}} \left|\sum\limits_{m=1}^M \eta_{mk}^{1/2}g_{mk}\hat{g}_{mk}^\ast\right|^2 }{ \Pd^{\mathrm{cf}}
    \sum\limits_{k'\neq k}^K \left|\sum\limits_{m=1}^M\eta_{mk'}^{1/2}g_{mk}\hat{g}_{mk'}^\ast\right|^2 +1}
    \!\right)},
\end{align}
which represents an achievable   rate for a genie-aided user that
knows the instantaneous channel gain.  Figure~\ref{fig:harden}
shows a comparison between \eqref{eq:Theo_rateexpr1}, which
assumes that the users only know the channel statistics, and the
genie-aided rate \eqref{eq aaa}, which assumes knowledge of the
realizations. As seen in the figure, the gap is small, which means
that downlink training is not necessary.
\end{remark}

\subsubsection{Achievable Uplink  Rate} \label{sec:ulrate}

The central processing unit detects the desired signal $q_k$ from
$r_{\ul,k}$ in \eqref{eq ULtrans
  21}. We assume that the central
  processing unit uses only statistical knowledge of the channel when
  performing the detection. Using a similar methodology as in
Section~\ref{sec DLrate}, we obtain a rigorous  closed-form
expression for the achievable uplink rate as follows.

\begin{theorem}\label{theorem ulrate}
An achievable uplink rate for the $k$th user in the Cell-Free
Massive MIMO system with matched filtering detection, for any $M$
and $K$, is given by \eqref{eq:Theo_uprateexpr1}, shown at the top of the page.
\end{theorem}

\begin{remark}
In the special case that all APs are collocated and all $K$ pilot
sequences are pairwisely orthogonal, then $\beta_{mk}
=\beta_{m'k}\triangleq \beta_{k}$, $\gamma_{mk}
=\gamma_{m'k}\triangleq \gamma_{k}$, and
$\pmb{\varphi}_{k}^H\pmb{\varphi}_{k'}=0, \forall k'\neq k$.
Equation \eqref{eq:Theo_uprateexpr1} then reduces to
\setcounter{equation}{27}\begin{align}\label{eq:Theo_uprateexpr2}
R_{\ul,k}^{\cf}
 =
 \log_2
    \left(
    1 + \frac{M\Pu^{\mathrm{cf}} \eta_k  \gamma_{k}  }{ \Pu^{\mathrm{cf}}\sum_{k'=1}^K\eta_{k'}\beta_{k'} +1 }
    \right),
\end{align}
\end{remark}
which is precisely the uplink capacity lower bound of a
single-cell Massive MIMO system with a collocated array obtained
in \cite{YM:13:ACCCC}, and a variation on that in
\cite{NLM:13:TCOM}.

\section{Pilot Assignment and Power Control} \label{sec PA-PC}

To obtain good system performance, the available radio resources
must be efficiently managed. In this section, we will present
methods for pilot sequence assignment and power control.
Importantly, pilot assignment and power control can be performed
independently, because the pilots are not power controlled.

\subsection{Greedy Pilot
    Assignment} \label{sec PiAssgn}
Typically, different users must use
  non-orthogonal pilot sequences, due to the limited length of the coherence
  interval. Since the length of the pilot sequences is $\taucf$, there
  exist $\taucf$ orthogonal pilot sequences. Here we focus on the case
  that $\taucf < K$. If $\taucf \geq K$, we simply assign $K$
  orthogonal pilot sequences to the $K$ users.

A simple baseline method for assigning
  pilot sequences of length $\taucf$ samples to the $K$ users is
  random pilot assignment \cite{AFMM:15:WCL}. With random pilot
  assignment, each user will be randomly assigned one pilot sequence
  from a predetermined set $\mathcal{S}_{\pmb{\varphi}}$ of $\taucf$
  orthogonal pilot sequences. Random pilot assignment could
  alternatively be done by letting each user choose an arbitrary
  unit-norm vector (i.e. not from a predetermined set of pilots).
  However, it appears from simulations that the latter scheme does not
  work well. While random pilot assignment is a useful baseline,
  occasionally two users in close vicinity of each other will use the
  same pilot sequence, which results in strong pilot contamination.

Optimal pilot assignment is a difficult combinatorial problem.  We
propose to use a simple greedy algorithm, which iteratively
refines the pilot assignment. The $K$ users are first randomly
assigned $K$ pilot sequences. Then the user that has the lowest
downlink rate, say user $k^\ast$, updates its pilot sequence so
that its pilot contamination effect is minimized.\footnote{In
principle, this ``worst
  user'' could be taken to be the user that has either the lowest
  uplink or the lowest downlink rate. In our numerical experiments, we
  reassign the pilot of the user having the lowest downlink rate,
  hence giving downlink performance some priority over uplink
  performance.}  The pilot contamination effect at the $k^\ast$th user
  is quantified by the second term in \eqref{eq:tidley1} which has
  variance
\begin{align}\label{eq GPA 1}
\EX{\left|\sum_{k'\neq k^\ast}^K g_{mk'} \pmb{\varphi}_{k^\ast}^H
\pmb{\varphi}_{k'} \right|^2} = \sum_{k'\neq k^\ast}^K \beta_{mk'}
\left|\pmb{\varphi}_{k^\ast}^H \pmb{\varphi}_{k'} \right|^2.
\end{align}
The $k^\ast$th user is assigned a new pilot sequence which
minimizes the pilot contamination in \eqref{eq GPA 1}, summed over
all APs:
 \begin{align}\label{eq GPA 2}
&\arg\min_{\pmb{\varphi}_{k^\ast}} \sum_{m=1}^M\sum_{k'\neq
k^\ast}^K \beta_{mk'} \left|\pmb{\varphi}_{k^\ast}^H
\pmb{\varphi}_{k'} \right|^2\nonumber\\
 &=
 \arg\min_{\pmb{\varphi}_{k^\ast}}   \frac{ \pmb{\varphi}_{k^\ast}^H \left(\sum_{m=1}^M\sum_{k'\neq k^\ast}^K
\beta_{mk'} \pmb{\varphi}_{k'}\pmb{\varphi}_{k'}^H\right)
\pmb{\varphi}_{k^\ast}}{\pmb{\varphi}_{k^\ast}^H\pmb{\varphi}_{k^\ast}},
\end{align}
where we used the fact that $\|\pmb{\varphi}_{k^\ast}\|^2=1$.  The
algorithm then proceeds iteratively for a predetermined number of
iterations.

The greedy pilot assignment algorithm can be summarized as
follows.

\hrulefill
\begin{algorithm}[Greedy pilot assignment]\label{sec: algogreedy}

~

\begin{itemize}
\item[1)] \emph{Initialization}: choose $K$ pilot sequences
$\pmb{\varphi}_{1}, \cdots, \pmb{\varphi}_{K}$ using the random
pilot assignment method. Choose the number of iterations, $N$, and
set $n=1$.

\item[2)] Compute $R_{\dl,k}^{\cf}$, using
\eqref{eq:Theo_rateexpr1}. Find the user with the lowest rate:
\begin{align}
k^\ast = \arg \min_k R_{\dl,k}^{\cf}.
\end{align}

\item[3)] Update the pilot sequence for the $k^\ast$th user by
choosing $\pmb{\varphi}_{k^\ast}$ from
$\mathcal{S}_{\pmb{\varphi}}$ which minimizes $$
\sum_{m=1}^M\sum_{k'\neq k^\ast}^K \beta_{mk'}
\left|\pmb{\varphi}_{k^\ast}^H \pmb{\varphi}_{k'} \right|^2.$$

\item[4)] Set $n := n+1$. Stop if $n>N$. Otherwise, go to Step 2.
\end{itemize}
\end{algorithm}
\hrulefill

\begin{remark}\label{remark_greedypilot}
The greedy pilot assignment can be performed at the CPU,  which
connects to all APs via backhaul links.  The pilot assignment is
recomputed on the large-scale fading time scale.\footnote{Hence
this
    recomputation is infrequent even in high mobility.  For example,
    at user mobility of $v=100$~km/h, and a carrier frequency of
    $f_{\text{c}}=2$ GHz, the channel coherence time is on the order
    of a millisecond.  The large-scale fading changes much more
    slowly, at least some $40$ times slower according to
    \cite{Rapp:96:Book,AML:IT:14}. As a result, the greedy pilot assignment method
    must only be done a few times per second.}   This simplifies the
signal processing at the central unit significantly. Furthermore,
since $\pmb{\varphi}_{k^\ast}$ is chosen from
$\mathcal{S}_{\pmb{\varphi}}$, to inform the users about their
assigned pilots, the CPU only needs to send an index to each user.
\end{remark}

\subsection{Power Control}\label{cellfree_powercontrol}

We next show that Cell-Free Massive MIMO can provide uniformly
good service to all users, regardless of their geographical
location, by using max-min power control.  While power control in
general is a well studied topic, the
  max-min power control problems that arise when optimizing Cell-Free
  Massive MIMO are entirely new. The power control is performed at the CPU,
   and importantly, is done on the large-scale fading
  time scale.

\subsubsection{Downlink}

In the downlink, given realizations of the large-scale fading, we
find the power control coefficients $\eta_{mk}$, $m=1, \cdots, M,
k=1, \cdots, K$, that maximize the minimum of the downlink rates
of all users, under the power constraint \eqref{eq:pct}.  At the
optimum point, all users get the same rate. Mathematically:
\begin{align}\label{eq opt 1}
    \left.%
\begin{array}{ll}
  \displaystyle\max_{\{\eta_{mk}\}} & \min\limits_{k=1, \cdots, K} R_{\dl,k}^{\cf}  \\
  \text{subject to}
     & \sum_{k=1}^K \eta_{mk}\gamma_{mk} \leq 1, ~  m=1, ..., M\\
     &  \eta_{mk} \geq 0, ~ k=1, ..., K, ~  m=1, ..., M,\\
\end{array}%
\right.
\end{align}
where $R_{\dl,k}^{\cf}$ is given by \eqref{eq:Theo_rateexpr1}.
Define $\varsigma_{mk}\triangleq \eta_{mk}^{1/2}$. Then, from
\eqref{eq:Theo_rateexpr1}, \eqref{eq opt 1} is equivalent to
\begin{align}\label{eq opt 1ne}
    \left.%
\begin{array}{ll}
  \displaystyle\max_{\{\eta_{mk}\}} & \min\limits_{k=1, \cdots, K}\!\!\!
        \frac{
            \left(\sum_{m=1}^M \gamma_{mk}\varsigma_{mk} \right)^2
            }{
            \sum\limits_{k'\neq k}^K\!\!\!\!\xi_{kk'}\! \left(\!\sum\limits_{m=1}^M \!\!\!\frac{\gamma_{mk'}\beta_{mk}\varsigma_{mk'}}{\beta_{mk'}} \!\!\right)^2\!\!+\!\! \sum\limits_{m=1}^M\!\!\!\beta_{mk} \!\!\!\sum\limits_{k'=1}^K \!\!\!\gamma_{mk'}\varsigma_{mk'}^2 \!+\!\frac{1}{\Pd^{\mathrm{cf}}}
            }  \\
  \text{s.t.}
     & \sum_{k=1}^K \eta_{mk}\gamma_{mk} \leq 1, ~  m=1, ..., M\\
     &  \eta_{mk} \geq 0, ~ k=1, ..., K, ~  m=1, ..., M,\\
\end{array}%
\right.
\end{align}
where $\xi_{kk'}\triangleq |\pmb{\varphi}_{k'}^H\pmb{\varphi}_{k}|^2$.

By introducing slack variables $\varrho_{k'k}$ and
$\vartheta_{m}$, we  reformulate \eqref{eq opt 1ne} as follows:
\begin{align}\label{eq opt 2}
    \left.%
\begin{array}{ll}
  \displaystyle\max_{\{\varsigma_{mk}, \varrho_{k'k},  \vartheta_{m}\}} & \min_{k=1, \cdots, K}
    \frac{ \left(\sum_{m=1}^M \gamma_{mk}\varsigma_{mk} \right)^2 }{ \sum\limits_{k'\neq k}^K\!\!|\pmb{\varphi}_{k'}^H\pmb{\varphi}_{k}|^2 \varrho_{k'k}^2+\!\! \sum\limits_{m=1}^M\!\!\beta_{mk} \vartheta_{m}^2 +\frac{1}{\Pd^{\mathrm{cf}}} }  \\
  \text{subject to}
     & \sum_{k'=1}^K \gamma_{mk'}\varsigma_{mk'}^2 \leq \vartheta_{m}^2 , ~ m=1, ..., M\\
     &  \sum_{m=1}^M \gamma_{mk'}\frac{\beta_{mk}}{\beta_{mk'}}\varsigma_{mk'} \leq \varrho_{k'k}, ~ \forall k'\neq k\\
     &  0 \leq \vartheta_{m} \leq 1, ~ m=1, ..., M\\
     &  \varsigma_{mk} \geq 0, ~ k=1, ..., K, ~ m=1, ..., M.\\
\end{array}%
\right.
\end{align}
The equivalence between \eqref{eq opt 1ne} and \eqref{eq opt 2}
follows directly from the fact that the first and second
constraints in \eqref{eq opt 2} hold with equality at the optimum.

\begin{proposition}\label{prop1}
The objective function of \eqref{eq opt 2} is quasi-concave, and
the problem \eqref{eq opt 2} is quasi-concave.
\begin{IEEEproof}
See Appendix~\ref{sec app propPC}.
\end{IEEEproof}
\end{proposition}
Consequently, \eqref{eq opt 2} can be solved efficiently by a
bisection search, in each step solving a sequence of convex
feasibility problem \cite{BV:04:Book}.  Specifically, the
following algorithm  solves \eqref{eq opt 2}.

\hrulefill
\begin{algorithm}[Bisection algorithm for solving \eqref{eq opt 2}]\label{sec: algo}

~

\begin{itemize}
\item[1)] \emph{Initialization}: choose the initial values of
$t_{\text{min}}$ and $t_{\text{max}}$, where $t_{\text{min}}$ and
$t_{\text{max}}$ define a range of relevant values of
 the objective function in \eqref{eq opt 2}. Choose a tolerance
$\epsilon > 0$.

\item[2)] Set $t:=\frac{t_{\text{min}} + t_{\text{max}}}{2}$.
Solve the following convex feasibility program:
\begin{align}\label{eq opt al1}
    \left\{%
\begin{array}{l}
     \hspace{0.0 cm} \left\|\B{v}_k\right\| \leq \frac{1}{\sqrt{t}}\!\!\sum\limits_{m=1}^M\!\! \gamma_{mk}\varsigma_{mk}, ~ k=1, ..., K,\\
     \hspace{0.0 cm} \sum\limits_{k'=1}^K \gamma_{mk'}\varsigma_{mk'}^2 \leq \vartheta_{m}^2 , ~ m=1, ..., M,\\
     \hspace{0.0 cm}  \sum\limits_{m=1}^M \gamma_{mk'}\frac{\beta_{mk}}{\beta_{mk'}}\varsigma_{mk'} \leq \varrho_{k'k}, ~ \forall k'\neq k,\\
     \hspace{0.0 cm}  0 \leq \vartheta_{m} \leq 1, ~ m=1, ..., M,\\
     \hspace{0.0 cm}  \varsigma_{mk} \geq 0, ~ k=1, ..., K, ~ m=1, ..., M,\\
\end{array}%
\right.
\end{align}
 where $\B{v}_k\triangleq \left[\B{v}_{k1}^T\B{I}_{-k} \quad
\B{v}_{k2}^T \quad \frac{1}{\sqrt{\Pd^{\mathrm{cf}}}}\right]^T$,
and where $\B{v}_{k1}\triangleq \left[
\pmb{\varphi}_{1}^H\pmb{\varphi}_{k}\varrho_{1k} ~ ... ~
\pmb{\varphi}_{K}^H\pmb{\varphi}_{K}\varrho_{Kk}\right]^T$,
$\B{I}_{-k}$ is a $K\times (K-1)$ matrix obtained from the
$K\times K$ identity matrix with the $k$th column removed, and
$\B{v}_{k2}\triangleq \left[\sqrt{\beta_{1k}}\vartheta_{1} ~ ... ~
\sqrt{\beta_{Mk}}\vartheta_{M}\right]^T$.

\item[3)] If problem \eqref{eq opt al1} is feasible, then set
$t_{\text{min}}:=t$, else set $t_{\text{max}}:=t$.

\item[4)] Stop if $t_{\text{max}}-t_{\text{min}} < \epsilon$.
Otherwise, go to Step 2.
\end{itemize}
\end{algorithm}
\hrulefill

\begin{remark}
The max-min power control problem can be directly extended to a
max-min weighted rate problem, where the $K$ users are weighted
according to priority: $\max \min\{w_kR_k\}$, where $w_k>0$ is the
weighting factor of the $k$th user. A user with higher priority
will be assigned a smaller weighting factor.
\end{remark}

\subsubsection{Uplink}
In the uplink, the max-min power control problem can be formulated
as follows:
\begin{align}\label{eq upopt 1}
    \left.%
\begin{array}{ll}
  \displaystyle\max_{\{\eta_{k}\}} & \min_{k=1, \cdots, K} R_{\ul,k}^{\cf}  \\
  \text{subject to}
     & 0 \leq \eta_{k} \leq 1, ~  k=1, ..., K,\\
\end{array}%
\right.
\end{align}
where $R_{\ul,k}^{\cf}$ is given by \eqref{eq:Theo_uprateexpr1}.
Problem \eqref{eq upopt 1} can be equivalently reformulated as
\begin{align}\label{eq upopt 2}
    \left.%
\begin{array}{ll}
  \displaystyle\max_{\{\eta_{k}\}, t} & t  \\
  \text{subject to}
     & t \leq R_{\ul,k}^{\cf}, ~  k=1, ..., K\\
     & 0 \leq \eta_{k} \leq 1, ~  k=1, ..., K.\\
\end{array}%
\right.
\end{align}

\begin{proposition}\label{prop2}
The optimization problem \eqref{eq upopt 2} is quasi-linear.
\begin{IEEEproof}
From \eqref{eq:Theo_uprateexpr1}, for a given $t$, all
inequalities involved in \eqref{eq upopt 2} are linear, and hence,
the program \eqref{eq upopt 2} is quasi-linear.
\end{IEEEproof}
\end{proposition}
Consequently, Problem \eqref{eq upopt 2} can be efficiently solved
by using bisection and solving a sequence of linear feasibility
problems.

\section{Small-Cell System} \label{sec small cell}

In this section, we give the system model, achievable rate
expressions, and max-min power control for small-cell systems.
These will be used in Section~\ref{sec numerical-result} where we
compare the performance of Cell-Free Massive MIMO and small-cell
systems.

For small-cell systems, we assume that each user is served by only
one AP. For each user, the available AP with the largest average
received useful signal power is selected. If an AP has already
been chosen by another user, this AP becomes unavailable. The AP
selection is done user by user in a random order. Let $m_k$ be the
AP chosen by the $k$th user. Then,
\begin{align}\label{eq sc choose}
m_k\triangleq
\arg\!\!\!\!\!\!\!\!\!\!\!\!\max_{m\in\{\text{available
APs}\}}\beta_{mk}.
\end{align}
We consider a short enough time scale
  that handovers between APs do not occur. This modeling choice was
  made to enable a rigorous performance analysis.   While there is
  precedent for this assumption in other literature
  \cite{LD:14:WCOM,ABG:11:TCOM}, future work may address the issue of
  handovers. As a result of this assumption,
  the performance figures we obtain
  for small-cell systems may be overoptimistic.

In contrast to Cell-Massive MIMO, in
  the small-cell systems, the channel does not harden. Specifically, while in Cell-Free Massive
  MIMO the effective channel is an inner product between two $M$-vectors---hence
  close to its mean, in the
  small-cell case the effective channel is a single Rayleigh fading scalar coefficient.
   Consequently, both
  the users and the APs must estimate their effective channel gain in
  order to demodulate the symbols, which requires both uplink and
  downlink training. The detailed transmission
  protocols for the uplink and downlink of small-cell systems are as
  follows.

\subsection{Downlink Transmission} \label{sec:SC_dl}

 In the downlink, the users
first estimate their channels based on  pilots sent from the APs.
The so-obtained channel estimates are used to detect the desired
signals.

Let $\tauscdl$ be the downlink training duration in samples,
$\sqrt{\tauscdl}\pmb{\phi}_{k} \in \mathbb{C}^{\tauscdl\times 1}$,
where $\|\pmb{\phi}_{k}\|^2=1$, is the pilot sequence transmitted
from the $m_k$th AP, and $\rho_{\mathrm{d,p}}^\sce$ is the
transmit power per downlink pilot symbol. The  MMSE estimate of
$g_{{m_k}k}$ can be expressed as
\begin{align}\label{eq DL CE1}
    \hat{g}_{{m_k}k} = g_{{m_k}k} - \varepsilon_{{m_k}k},
\end{align}
where $\varepsilon_{{m_k}k}$ is the channel estimation error,
which is independent of the channel estimate $\hat{g}_{{m_k}k}$.
Furthermore, we have $\hat{g}_{{m_k}k}\sim \CG{0}{ \mu_{{m_k}k} }$
and $\varepsilon_{{m_k}k}\sim
\CG{0}{\beta_{{m_k}k}-\mu_{{m_k}k}}$, where
\begin{align}\label{eq DL CE2}
    \mu_{{m_k}k}
        \triangleq
        \frac{\tauscdl\rho_{\mathrm{d,p}}^\sce\beta_{{m_k}k}^2}{\tauscdl\rho_{\mathrm{d,p}}^\sce\sum_{k'=1}^K\beta_{{m_{k'}}k}\left|\pmb{\phi}_k^H \pmb{\phi}_{k'}\right|^2+1}.
\end{align}

After sending the pilots for the channel estimation, the $K$
chosen APs send the data. Let $\sqrt{\alpha_{\mathrm{d},k}}q_k$,
$\EX{|q_k|^2}=1$, be the symbol transmitted from the $m_k$th AP,
destined for the $k$th user, where $\alpha_{\mathrm{d},k}$ is a
power control coefficient, $0\leq \alpha_{\mathrm{d},k} \leq 1$.
The $k$th user receives
\begin{align}
    y_k
    &= \sqrt{\Pd^\sce}\sum_{k'=1}^K g_{{m_{k'}}k} \sqrt{\alpha_{\mathrm{d},k'}}
    q_{k'} + w_k\nonumber\\ \label{eq scDL 1b}
    &=\sqrt{\Pd^\sce} \hat{g}_{{m_{k}}k} \sqrt{\alpha_{\mathrm{d},k}} q_{k} + \sqrt{\Pd^\sce} \varepsilon_{{m_{k}}k} \sqrt{\alpha_{\mathrm{d},k}} q_{k} \nonumber\\
    &+ \sqrt{\Pd^\sce}\sum_{k'\neq k}^K g_{{m_{k'}}k} \sqrt{\alpha_{\mathrm{d},k'}}
    q_{k'} + w_k,
\end{align}
where $\Pd^\sce$ is the normalized downlink transmit SNR and $w_k
\sim \CG{0}{1}$ is additive Gaussian noise.

\begin{remark}\label{remark:sc_harden}
In small-cell systems, since only one
  single-antenna AP is involved in transmission to a given user, the
  concept of ``conjugate beamforming'' becomes void. Downlink
  transmission entails only transmitting the symbol destined for the
  $k$th user, appropriately scaled to meet the transmit power
  constraint.  Channel estimation at the user is required in order to
  demodulate, as  there is no channel hardening (see discussion above).
\end{remark}

\subsubsection{Achievable Downlink  Rate}

Treating the last three terms of \eqref{eq scDL 1b} as
uncorrelated effective noise, we obtain the achievable
downlink rate for the $k$th user as in \eqref{eq RateSC1a}, shown at the top of the page.
\setcounter{eqnback}{\value{equation}} \setcounter{equation}{41}
\begin{figure*}[!t]
\begin{align}\label{eq RateSC1a}
    R_{\dl,k}^{\sce}
    &=
    \EX{\log_2
        \left(
        1+ \frac{\Pd^\sce\alpha_{\mathrm{d},k}\left|\hat{g}_{{m_k}k}\right|^2}{\Pd^\sce\alpha_{\mathrm{d},k}(\beta_{m_kk} - \mu_{m_kk})+\Pd^\sce\sum\limits_{k'\neq k}^K\alpha_{\mathrm{d},k'}\beta_{{m_{k'}}k} + 1}
        \right)},
\end{align}
\hrulefill
\end{figure*}
\setcounter{eqncnt}{\value{equation}}
\setcounter{equation}{\value{eqnback}}

Since the channel does not harden, applying the bounding
techniques in Section~\ref{sec rate}, while not impossible in
principle, would yield very pessimistic capacity bounds. However,
since $\left|\hat{g}_{{m_k}k}\right|^2$ is exponentially
distributed with mean $\mu_{m_kk}$, the achievable rate in
\eqref{eq RateSC1a} can be expressed in  closed form in terms of
the exponential integral function $\text{Ei}(\cdot)$
\cite[Eq.~(8.211.1)]{GR:07:Book} as:
\setcounter{equation}{42}\begin{align}\label{eq RateSC 2}
     R_{\dl,k}^{\sce}
    &=
    -(\log_2e) e^{1/\bar{\mu}_{m_kk}} \text{Ei}\left(-\frac{1}{\bar{\mu}_{m_kk}}\right),
\end{align}
where
\begin{align}\label{eq RateSC 2b}
\bar{\mu}_{m_kk}\triangleq
\frac{\Pd^\sce\alpha_{\mathrm{d},k}{\mu}_{m_kk}}{\Pd^\sce\alpha_{\mathrm{d},k}(\beta_{m_kk}
- \mu_{m_kk})+\Pd^\sce\sum\limits_{k'\neq k}^K
\alpha_{\mathrm{d},k'}\beta_{{m_{k'}}k} + 1}.
\end{align}

\subsubsection{Max-Min Power Control}

As in the Cell-Free Massive MIMO systems, we consider max-min
power control which can be formulated as follows:
\begin{align}\label{eq dlsc-opt 1}
    \left.%
\begin{array}{ll}
  \displaystyle\max_{\{\alpha_{\mathrm{d},k}\}}& \min_{k=1, \cdots, K} R_{\dl,k}^{\sce}  \\
  \text{subject to}
     & 0 \leq \alpha_{\mathrm{d},k} \leq 1, ~  k=1, \cdots, K.\\
\end{array}%
\right.
\end{align}
Since $R_{\text{d},k}^{\mathrm{sc}}$ is a monotonically increasing
function of $\bar{\mu}_{m_kk}$, \eqref{eq dlsc-opt 1} is
equivalent to
\begin{align}\label{eq dlsc-opt 2}
    \left.%
\begin{array}{ll}
  \displaystyle\max_{\{\alpha_{\mathrm{d},k}\}} & \min_{k=1, \cdots, K}  \bar{\mu}_{m_kk} \\
  \text{subject to}
     & 0 \leq \alpha_{\mathrm{d},k} \leq 1, ~  k=1, \cdots, K.\\
\end{array}%
\right.
\end{align}
Problem \eqref{eq dlsc-opt 2} is a quasi-linear program, which can
be solved by using bisection.

\subsection{Uplink Transmission}

In the uplink, the APs first estimate the channels based on pilots
sent from the users. The so-obtained channel estimates are used to
detect the desired signals. Let $\Pu^\sce$ and
$0\leq\alpha_{\mathrm{u},k} \leq 1$ be the normalized   SNR and
the power control coefficient at the $k$th user, respectively.
Then, following the same methodology as in the derivation of the
downlink transmission, we obtain the following achievable uplink
rate for the $k$th user:
\begin{align}\label{eq RateSC 22}
     R_{\ul,k}^{\sce}
    &=
    -(\log_2e) e^{1/\bar{\omega}_{m_kk}} \text{Ei}\left(-\frac{1}{\bar{\omega}_{m_kk}}\right),
\end{align}
where
\begin{align}\label{eq RateSC 2b2}
\bar{\omega}_{m_kk}\triangleq
\frac{\Pu^\sce\alpha_{\mathrm{u},k}{\omega}_{m_kk}}{\Pu^\sce\alpha_{\mathrm{u},k}(\beta_{m_kk}
- \omega_{m_kk})+\Pu^\sce\sum\limits_{k'\neq k}^K
\alpha_{\mathrm{u},k'}\beta_{{m_{k}}k'} + 1},
\end{align}
and where $\omega_{m_kk}$ is given by
\begin{align}\label{eq ULSC CE1}
    \omega_{{m_k}k}
        \triangleq
        \frac{\tauscul\rho_{\mathrm{u,p}}^\sce\beta_{{m_k}k}^2}{\tauscul\rho_{\mathrm{u,p}}^\sce\sum_{k'=1}^K\beta_{{m_{k}}k'}\left|\pmb{\psi}_k^H \pmb{\psi}_{k'}\right|^2+1}.
\end{align}
In \eqref{eq ULSC CE1}, $\tauscul$ is the uplink training duration
in samples, $\sqrt{\tauscul}\pmb{\psi}_k\in
\mathbb{C}^{\tauscul\times 1}$, where $\|\pmb{\psi}_k\|^2=1$, is
the pilot sequence transmitted from the $k$th user, and
$\rho_{\mathrm{u,p}}^\sce$ is the transmit power per uplink pilot
symbol.

Similarly to in the downlink, the max-min power control problem
for the uplink can be formulated as a quasi-linear program:
\begin{align}\label{eq ulsc-opt 2}
    \left.%
\begin{array}{ll}
  \displaystyle\max_{\{\alpha_{\mathrm{u},k}\}}& \min_{k=1, \cdots, K}  \bar{\omega}_{m_kk} \\
  \text{subject to}
     & 0 \leq \alpha_{\mathrm{u},k} \leq 1, ~  k=1, \cdots, K,\\
\end{array}%
\right.
\end{align}
which can be solved by using   bisection.

\section{Numerical Results and Discussions}\label{sec numerical-result}

We quantitatively study the performance of Cell-Free Massive MIMO,
and compare it to that of small-cell systems. We specifically
demonstrate the effects of shadow fading correlation. The $M$ APs
and $K$ users are uniformly distributed at random within a square
of size $D\times D$ $\text{km}^2$.

\subsection{Large-Scale Fading Model}

We describe the path loss and shadow fading correlation models,
which are used in the performance evaluation. The large-scale
fading coefficient $\beta_{mk}$ in \eqref{eq:gmk} models the path
loss and shadow fading, according to
\begin{align}\label{eq:betmk}
\beta_{mk} = \text{PL}_{mk}\cdot
10^{\frac{\sigma_{\text{sh}}z_{mk}}{10}},
\end{align}
where $\text{PL}_{mk}$ represents the path loss, and
$10^{\frac{\sigma_{\text{sh}}z_{mk}}{10}}$ represents the shadow
fading with the standard deviation $\sigma_{\text{sh}}$, and
$z_{mk}\sim\mathcal{N}(0,1)$.

\subsubsection{Path loss Model} \label{sec_pathloss}

We use a three-slope model for the path loss \cite{TSG:01:VTC}:
the path loss exponent equals $3.5$ if distance between the $m$th
AP and the $k$th user (denoted by $d_{mk}$) is greater than $d_1$,
equals $2$ if $d_1\geq d_{mk}> d_0$, and equals $0$ if $d_{mk}
\leq d_0$ for some $d_0$ and $d_1$. When $d_{mk}>d_1$, we employ
the Hata-COST231 propagation model. More precisely, the path loss
in dB is given by {\small\begin{align}\label{eq:ploss}
\text{PL}_{mk} \!=\! \left\{\!
\begin{array}{l}
  -L - 35\log_{10} (d_{mk}), ~ \text{if} ~ d_{mk}>d_1\\
  -L - 15\log_{10} (d_1) - 20\log_{10} (d_{mk}), ~ \text{if} ~ d_0< \!d_{mk}\leq d_1\\
  -L - 15\log_{10} (d_1) - 20\log_{10} (d_{0}), ~ \text{if} ~ d_{mk} \leq d_0\\
\end{array}%
\right.
 \end{align}}where
\begin{align}\label{eq:ploss1}
L&\triangleq 46.3+33.9\log_{10}(f)-13.82\log_{10}(h_{\text{AP}})\nonumber\\
&-
(1.1\log_{10}(f)-0.7)h_{\text{u}}+(1.56\log_{10}(f)-0.8),
\end{align}
and where $f$ is the carrier frequency (in MHz), $h_{\text{AP}}$
is the AP antenna height (in m), and $h_{\text{u}}$ denotes the
user antenna height (in m). The path loss $\text{PL}_{mk}$ is a
continuous function of $d_{mk}$. Note that when $d_{mk} \leq d_1$,
there is no shadowing.

\subsubsection{Shadowing
Correlation Model} \label{sec ShaCorrelation}

Most previous work assumed that the shadowing coefficients (and
therefore $z_{mk}$) are uncorrelated. However, in practice,
transmitters/receivers that are in close vicinity of each other
may be surrounded by common obstacles, and hence, the shadowing
coefficients are correlated. This correlation may significantly
affect the system performance.

For the shadow fading coefficients, we
  will use a model with two components \cite{WTN:08:VT}:
\begin{align}\label{eq:correlation1}
z_{mk} =\sqrt{\delta}a_m + \sqrt{1-\delta}b_k, ~ m=1,\ldots, M, ~
K=1,\ldots, K,
\end{align}
where $a_m\sim\mathcal{N}(0,1)$ and $b_k\sim\mathcal{N}(0,1)$ are
independent random variables, and $\delta$, $0\leq\delta\leq 1$,
is a parameter.  The variable $a_m$ models contributions to the
shadow fading that result from obstructing objects in the vicinity
of the $m$th AP, and which affects the channel from that AP to all
users in the same way. The variable $b_k$ models contributions to
the shadow fading that result from objects in the vicinity of the
$k$th user, and which affects the channels from that user to all
APs in the same way.  When $\delta=0$, the shadow fading from a
given user is the same to all APs, but different users are
affected by different shadow fading. Conversely, when $\delta=1$,
the shadow fading from a given AP is the same to all users;
however, different APs are affected by different shadow fading.
Varying $\delta$ between $0$ and $1$ trades off between these two
extremes.

The covariance functions of $a_m$ and
  $b_k$ are given by:
\begin{align}\label{eq: spatialCorr1}
\EX{a_ma_{m'}}
    =
    2^{-\frac{d_\mathrm{a}(m,m')}{d_{\mathrm{decorr}}}}, \quad \EX{b_kb_{k'}}
    =
    2^{-\frac{d_\mathrm{u}(k,k')  }{d_{\mathrm{decorr}}}},
\end{align}
where $d_\mathrm{a}(m,m')$ is the geographical distance between
the $m$th and $m'$th APs, $d_\mathrm{u}(k,k')$ is the geographical
distance between the $k$th and $k'$th users, and
$d_{\mathrm{decorr}}$ is a decorrelation distance which depends on
the environment. Typically, the decorrelation distance is on the
order of $20$--$200$ m. A shorter decorrelation distance
corresponds to an environment with a lower degree of stationarity.
This model for correlation between different geographical
locations has been validated both in theory and by practical
experiments \cite{Gud:91:EL,WTN:08:VT}.

\subsection{Parameters and Setup}

\begin{table}[b!]
    \caption{
        System parameters for the simulation.
    }
    \centerline{
\begin{tabular}{|c|c|}
  \hline
  Parameter & Value \\
  \hline
  Carrier frequency & $1.9$ GHz \\
    \hline
  Bandwidth & $20$ MHz \\
    \hline
  Noise figure (uplink and downlink)& $9$ dB \\
    \hline
 AP antenna height & $15$ m \\
  \hline
 User antenna height & $1.65$ m \\
  \hline
 $\bar{\rho}_{\mathrm{d}}^{\mathrm{cf}}$, $\bar{\rho}_{\mathrm{u}}^{\mathrm{cf}}$, $\bar{\rho}_{\mathrm{p}}^{\mathrm{cf}}$ & $200$, $100$, $100$ mW \\
  \hline
 $\sigma_{\text{sh}}$ & $8$ dB \\
  \hline
 $D$, $d_1$, $d_0$ & $1000$, $50$, $10$ m \\
  \hline
\end{tabular}}
    \label{table:1}
\end{table}

In all examples, we choose the parameters summarized in
Table~\ref{table:1}. The quantities
$\bar{\rho}_{\mathrm{d}}^{\mathrm{cf}}$,
$\bar{\rho}_{\mathrm{u}}^{\mathrm{cf}}$, and
$\bar{\rho}_{\mathrm{p}}^{\mathrm{cf}}$ in this table are the
transmit powers of downlink data, uplink data, and pilot symbols,
respectively. The corresponding normalized transmit SNRs
$\Pd^{\mathrm{cf}}$, $\Pu^{\mathrm{cf}}$, and $\Pp^{\mathrm{cf}}$
can be computed by dividing these powers by the noise power, where
the noise power is given by
$$\text{noise power} = \text{bandwidth}\times k_{\mathrm{B}}\times T_0 \times \text{noise
figure} ~\text{(W)},$$ where $k_{\mathrm{B}}=1.381\times 10^{-23}$
(Joule per Kelvin) is  the Boltzmann constant, and $T_0 = 290$
(Kelvin) is the noise temperature. To avoid boundary effects, and
to imitate a network with an infinite area, the square area is
wrapped around at the edges, and hence, the simulation area has
eight neighbors.

\begin{figure}[t!]
\centerline{\includegraphics[width=0.5\textwidth]{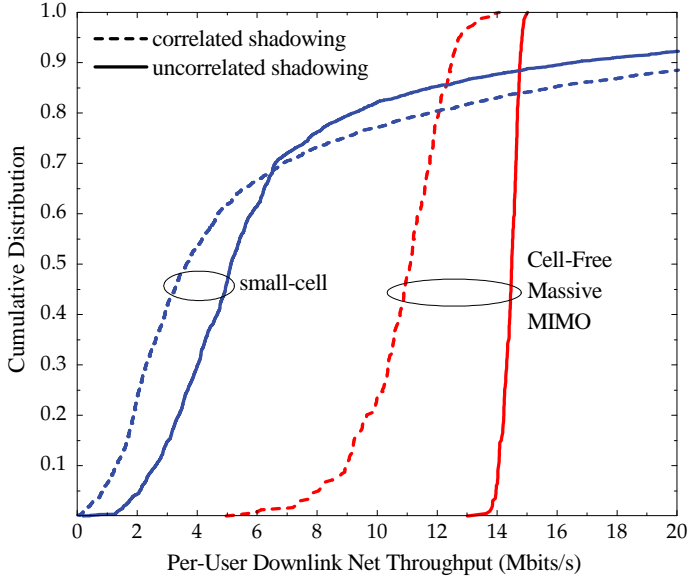}}
\caption{The cumulative distribution of the per-user downlink net
  throughput for correlated and uncorrelated shadow fading, with the
  greedy pilot assignment and max-min power control. Here, $M=100$,
  $K=40$, and $\tau^{\mathrm{cf}}=\tau^\sce_{\mathrm{d}}=20$.
\label{fig:2}}
\end{figure}

\begin{figure}[t!]
\centerline{\includegraphics[width=0.5\textwidth]{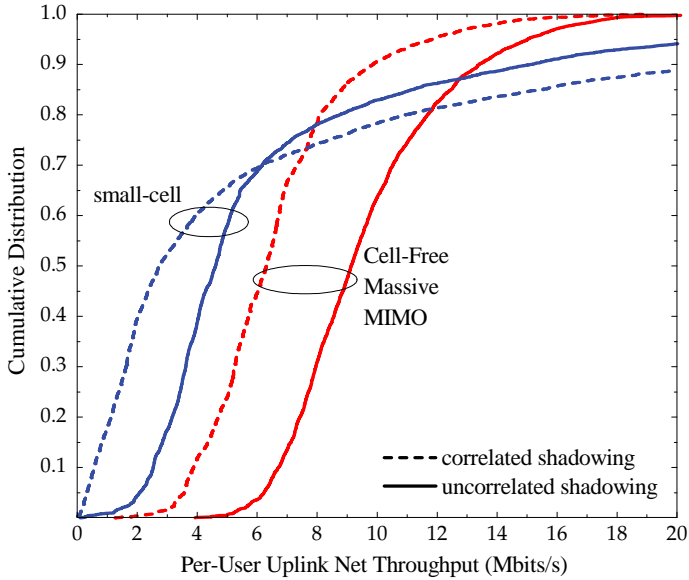}}
\caption{The same as Figure~\ref{fig:2} but for the uplink, and
  $\tau^{\mathrm{cf}}=\tau^\sce_{\mathrm{u}}=20$. \label{fig:3}}
\end{figure}

\begin{figure}[t!]
\centerline{\includegraphics[width=0.5\textwidth]{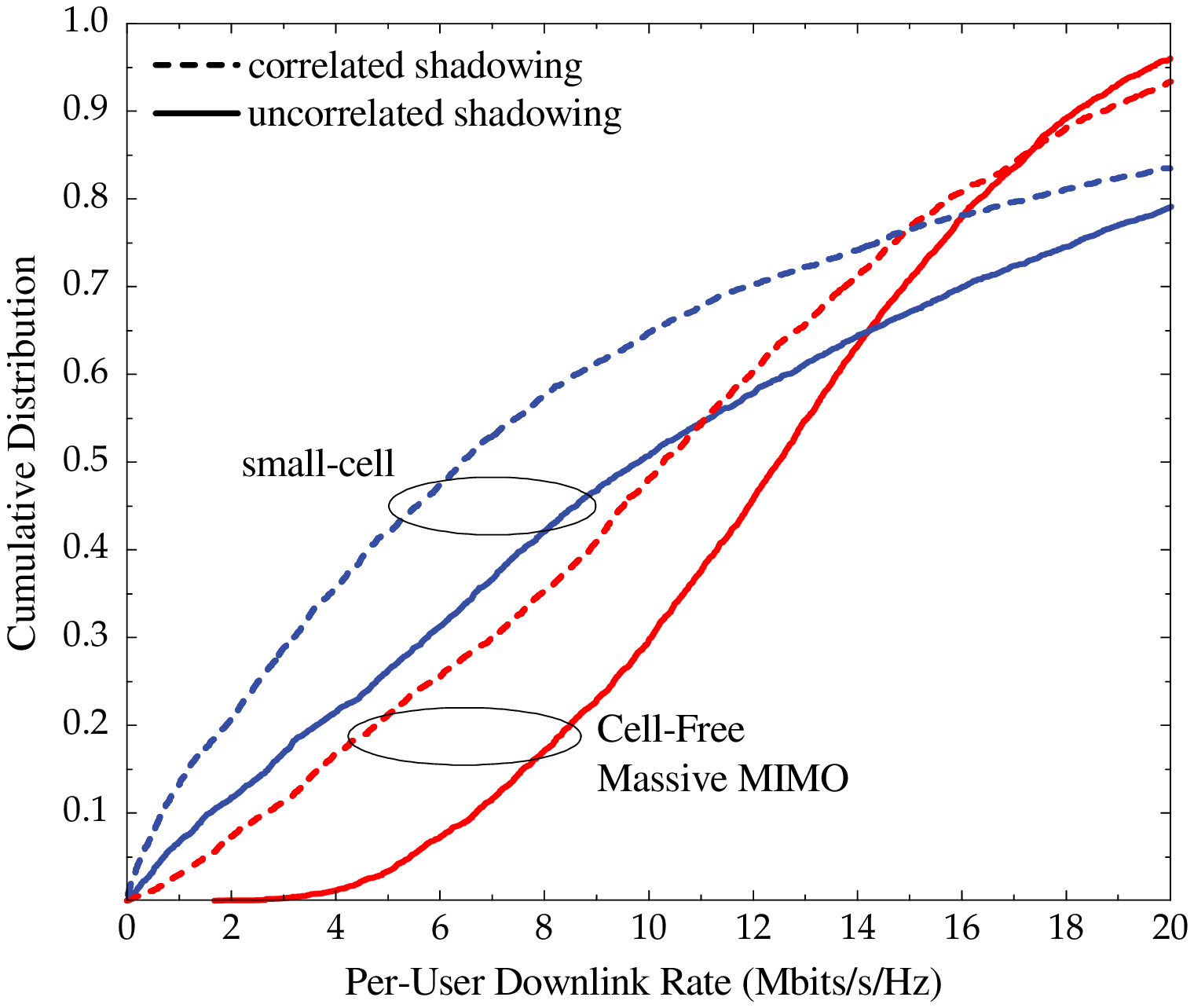}}
\caption{The cumulative distribution of the per-user downlink net
  throughput for correlated and uncorrelated shadow fading, with the
  greedy pilot assignment and without power control. Here, $M=100$,
  $K=40$, and $\tau^{\mathrm{cf}}=\tau^\sce_{\mathrm{d}}=20$.
\label{fig:4}}
\end{figure}

\begin{figure}[t!]
\centerline{\includegraphics[width=0.5\textwidth]{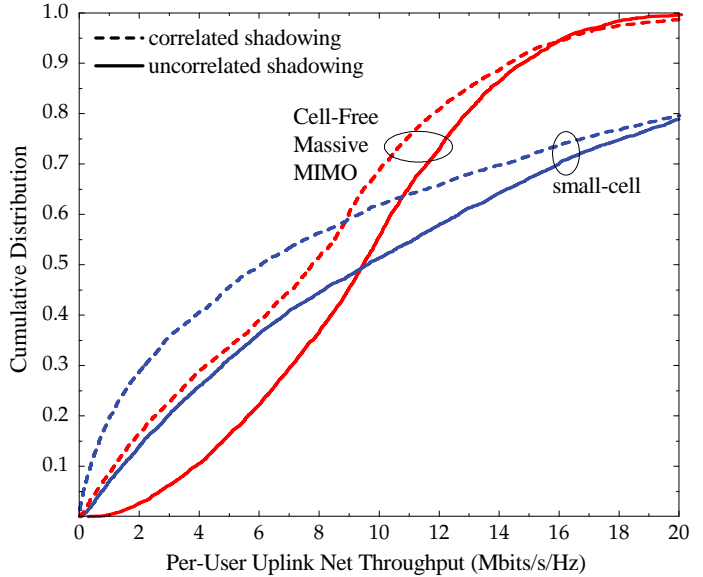}}
\caption{The same as Figure~\ref{fig:4} but for the uplink, and
  $\tau^{\mathrm{cf}}=\tau^\sce_{\mathrm{u}}=20$.\label{fig:5}}
\end{figure}

We consider the per-user net throughputs which take into account
the channel estimation overhead, and are defined as follows:
\begin{align} \label{eq spect1}
S_{{\tt A},k}^{\cf} &=
    B\frac{1-\tau^\cf/\tauc}{2}R_{{\tt A},k}^{\cf},\\ \label{eq spect2}
S_{{\tt A},k}^{\sce} &=
    B\frac{1-(\tauscdl+\tauscul)/\tauc}{2}R_{{\tt A},k}^{\sce},
\end{align}
where ${\tt A} \in \{\text{d}, \text{u}\}$ correspond to downlink
respectively uplink transmission, $B$ is the spectral bandwidth,
and $\tauc$ is again the coherence interval in samples. The terms
$\tau^\cf/\tauc$ and $(\tauscdl+\tauscul)/\tauc$ in \eqref{eq
spect1} and \eqref{eq spect2} reflect the fact that, for each
coherence interval of length $\tauc$ samples, in the Cell-Free
Massive MIMO systems, we spend $\tau^\cf$ samples for the uplink
training, while in the small-cell systems, we spend
$\tauscdl+\tauscul$ samples for the uplink and downlink training.
In all examples, we take $\tauc=200$ samples, corresponding to a
coherence bandwidth of $200$ KHz and a coherence time of $1$ ms,
and choose $B=20$ MHz.

To ensure a fair comparison between Cell-Free Massive MIMO and
small-cell systems, we choose $\Pd^\sce =
\frac{M}{K}\Pd^{\mathrm{cf}}$, $\Pu^\sce=\Pu^{\mathrm{cf}}$, and
$\rho_{\mathrm{u,p}}^\sce=\rho_{\mathrm{d,p}}^\sce =
\Pp^{\mathrm{cf}}$, which makes the total radiated power equal in
all cases. The cumulative distributions of the per-user
downlink/uplink net throughput in our examples are generated as
follows:
\begin{itemize}
    \item For the case with max-min power control: 1) $200$ random
      realizations of the AP/user locations and shadow fading profiles
      are generated; 2) for each realization, the per-user net
      throughputs of $K$ users are computed by using max-min power
      control as discussed in Section~\ref{cellfree_powercontrol} for
      Cell-Free Massive MIMO and in Section~\ref{sec small cell} for
      small-cell systems---with max-min power control these
      throughputs are the same for all users; 3) a cumulative
      distribution is generated over the so-obtained per-user net
      throughputs.

    \item For the case without power control: same procedure, but in
      2) no power control is performed.  Without power control, for
      Cell-Free Massive MIMO, in the downlink transmission, all APs
      transmit with full power, and at the $m$th AP, the power control
      coefficients $\eta_{mk}$, $k=1,\ldots K$, are the same, i.e.,
      $\eta_{mk} = \left(\sum_{k'=1}^K\gamma_{mk'} \right)^{-1}$,
      $\forall k=1, \ldots K$, (this
        directly comes from \eqref{eq:pct}), while in the uplink, all
      users transmit with full power, i.e., $\eta_k=1$, $\forall k=1,
      \ldots K$. For the small-cell system, in the downlink, all
      chosen APs transmit with full power,
      i.e. $\alpha_{\mathrm{d},k}=1$, and in the uplink, all users
      transmit with full power, i.e. $\alpha_{\mathrm{u},k}=1$, $k=1,
      \ldots K$.

    \item For the correlated shadow fading scenario, we use the
      shadowing correlation model discussed in Section~\ref{sec
        ShaCorrelation}, and we choose $d_{\mathrm{decorr}}=0.1$ km
      and $\delta=0.5$.

    \item For the small-cell systems, the greedy pilot assignment
      works in the same way as the scheme for Cell-Free Massive MIMO
      discussed in Section~\ref{sec PiAssgn}, except for that in the
      small-cell systems, since the chosen APs do not cooperate, the
      worst user will find a new pilot which minimizes the pilot
      contamination corresponding to its AP (rather than summed over
      all APs as in the case of Cell-Free systems).
\end{itemize}

\begin{figure}[t!]
\centerline{\includegraphics[width=0.5\textwidth]{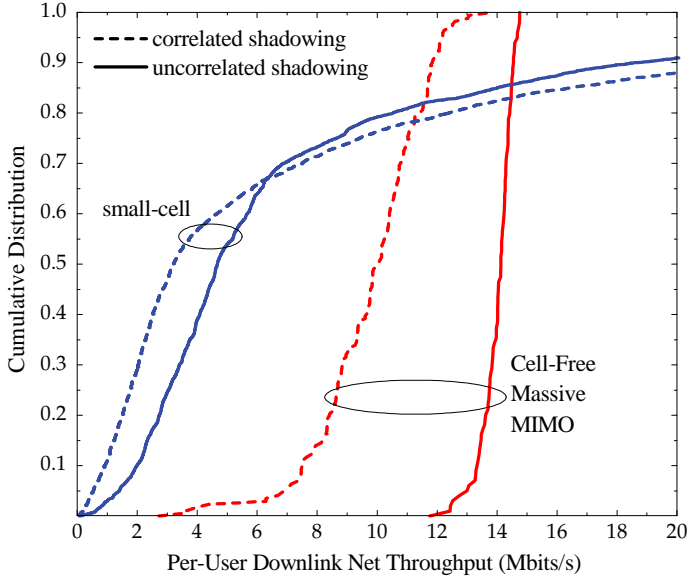}}
\caption{The cumulative distribution of the per-user downlink net
throughput for correlated and uncorrelated shadow fading, with the
random pilot assignment and max-min power control. Here, $M=100$,
$K=40$, and $\tau^{\mathrm{cf}}=\tau^\sce_{\mathrm{d}}=20$.
\label{fig:6}}
\end{figure}

\begin{figure}[t!]
\centerline{\includegraphics[width=0.5\textwidth]{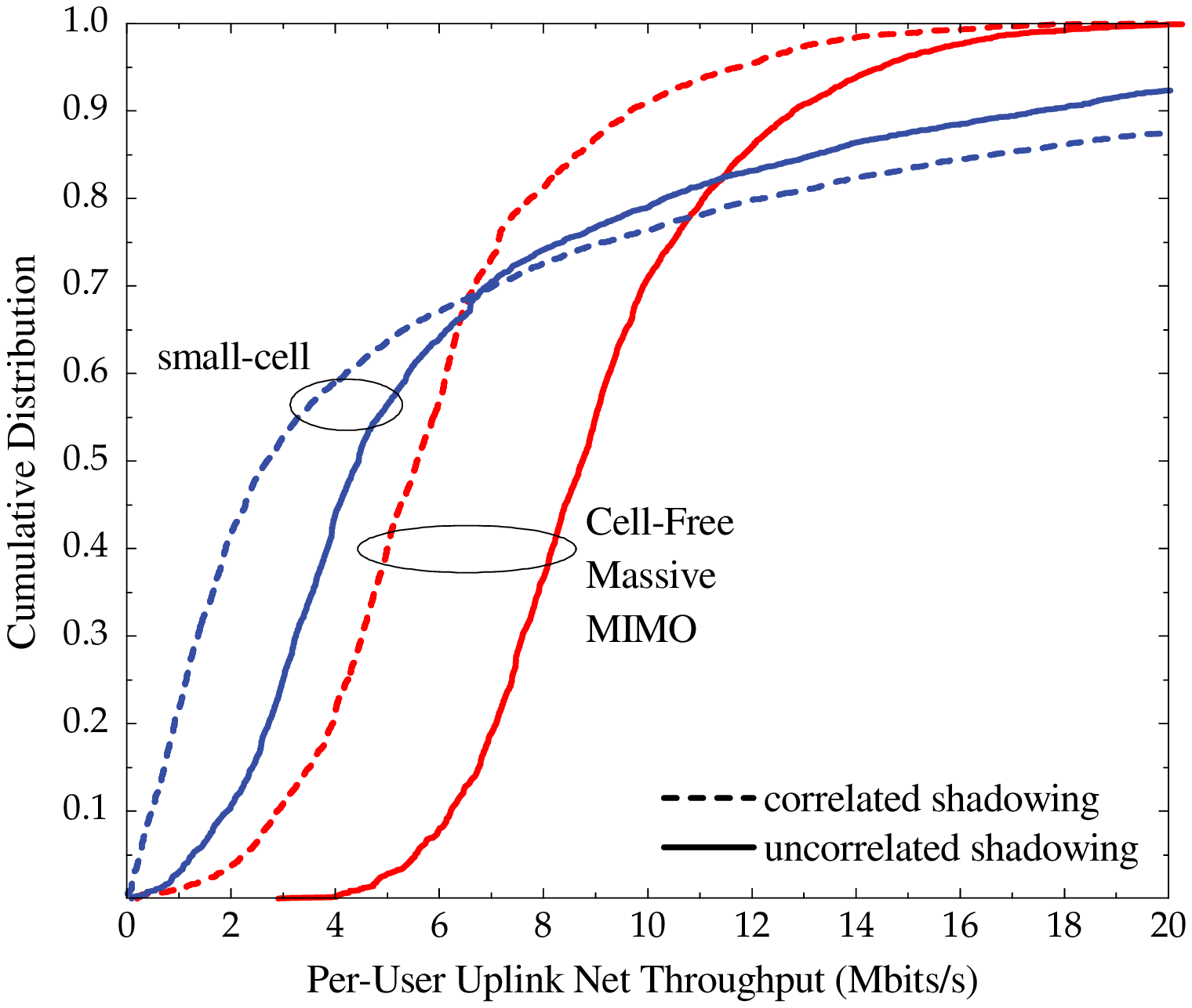}}
\caption{The same as Figure~\ref{fig:6} but for the uplink, and
$\tau^{\mathrm{cf}}=\tau^\sce_{\mathrm{u}}=20$. \label{fig:7}}
\end{figure}

\begin{figure}[t!]
\centerline{\includegraphics[width=0.5\textwidth]{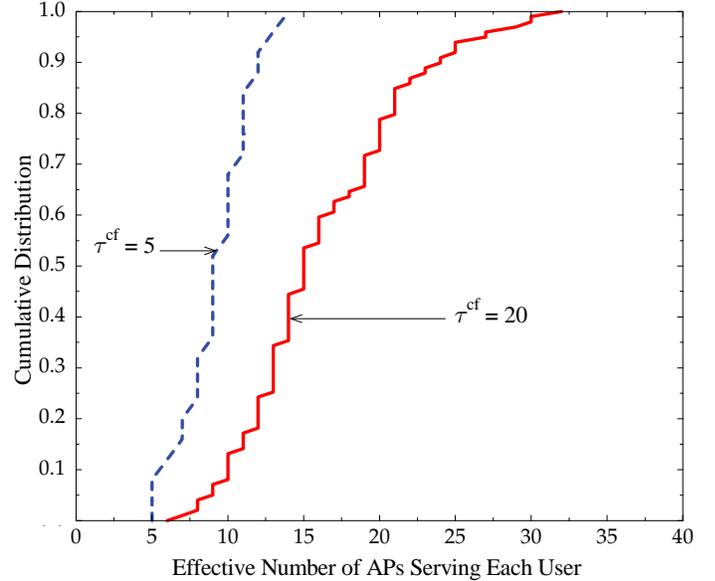}}
\caption{Cumulative distribution of the effective number of APs
serving each user. Here, $M=100$, $K=40$, and $\tau^\cf=5$ and
$20$.\label{fig:8}}
\end{figure}

\subsection{Results and Discussions}\label{sec:rnd}

We first compare the performance of Cell-Free Massive MIMO with
that of small-cell systems with greedy pilot assignment and
max-min power control. Figure~\ref{fig:2} compares the cumulative
distribution of the per-user downlink net throughput for Cell-Free
Massive MIMO and small-cell systems, with $M=100$, $K=40$, and
$\tau^\cf=\tauscdl=\tauscul=20$, with and without shadow fading
correlation.

Cell-Free Massive MIMO significantly outperforms small-cell in
both median and in 95\%-likely performance. The net throughput of
Cell-Free Massive MIMO  is much more concentrated around its
median, compared with the small-cell systems. Without shadow
fading correlation, the 95\%-likely net throughput of the
Cell-Free downlink is about $14$ Mbits/s which is $7$ times higher
than that of the small-cell downlink (about $2.1$ Mbits/s). In
particular, we can see that the small-cell systems are much more
affected by shadow fading correlation than Cell-Free Massive MIMO
is. This is due to the fact that when the shadowing coefficients
are highly correlated, the gain from choosing the best APs in a
small-cell system is reduced. With shadowing correlation, the
95\%-likely net throughput of the Cell-Free downlink is about $10$
times higher than that of the small-cell system. The same insights
can be obtained for the uplink, see Figure~\ref{fig:3}. In
addition, owing to the fact that the downlink uses more power
(since $M>K$ and $\Pd^{\mathrm{cf}} > \Pu^{\mathrm{cf}}$) and has
more power control coefficients to choose than the uplink does,
the downlink performance is better than the uplink performance.

Next we compare Cell-Free Massive MIMO and small-cell systems,
assuming that no power control is performed. Figures~\ref{fig:4}
and \ref{fig:5} show the cumulative distributions of the per-user
net throughput for the downlink and the uplink, respectively, with
$M=100$, $K=40$, and $\tau^\cf=\tauscdl=\tauscul=20$, and with the
greedy pilot assignment method. In both uncorrelated and
correlated shadowing scenarios, Cell-Free Massive MIMO outperforms
the small-cell approach in terms of 95\%-likely per-user net
throughput. In addition, a comparison of Figure~\ref{fig:2} (or
\ref{fig:3}) and Figure~\ref{fig:4} (or \ref{fig:5}) shows that
with power control, the performance of Cell-Free Massive MIMO
improves significantly in terms of both median and 95\%-likely
throughput.  In the uncorrelated shadow fading scenario, the power
allocation can improve the 95\%-likely Cell-Free throughput by a
factor of $2.5$ for the downlink and a factor of $2.3$ for the
uplink, compared with the case   without power control. For the
small-cell system, power control improves the 95\%-likely
throughput but not the median throughput (recall that the power
control policy explicitly aims at improving the performance of the
worst user).

In Figures~\ref{fig:6} and \ref{fig:7}, we consider the same
setting as in Figures~\ref{fig:2} and \ref{fig:3}, but here we use
the random pilot assignment scheme. These figures provide the same
insights as Figures~\ref{fig:2} and \ref{fig:3}. Furthermore, by
comparing these figures with Figures~\ref{fig:2} and \ref{fig:3},
we can see that with greedy pilot assignment, the 95\%-likely net
throughputs can be improved by about $20\%$ compared with when
random pilot assignment is used.

\begin{figure}[t!]
\centerline{\includegraphics[width=0.5\textwidth]{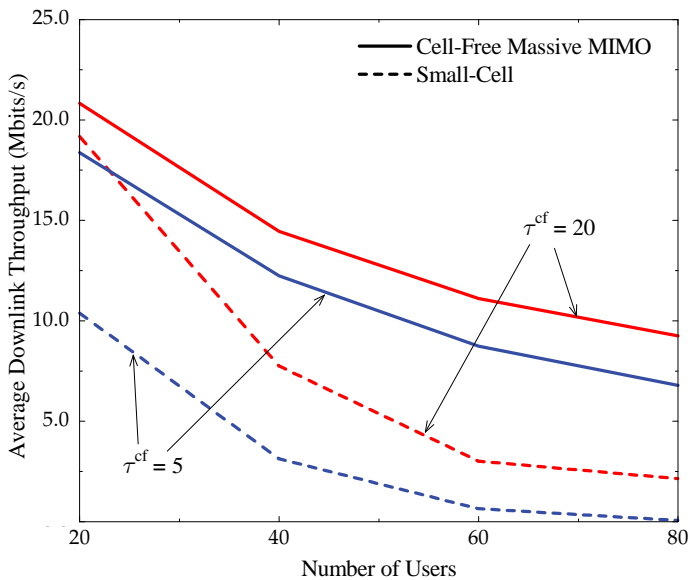}}
\caption{The average downlink net throughput versus the number
of users for different $\tau^\cf$. Here,
$M=100$.\label{fig:cellfree_smallcell_dl_K}}
\end{figure}

\begin{figure}[t!]
\centerline{\includegraphics[width=0.5\textwidth]{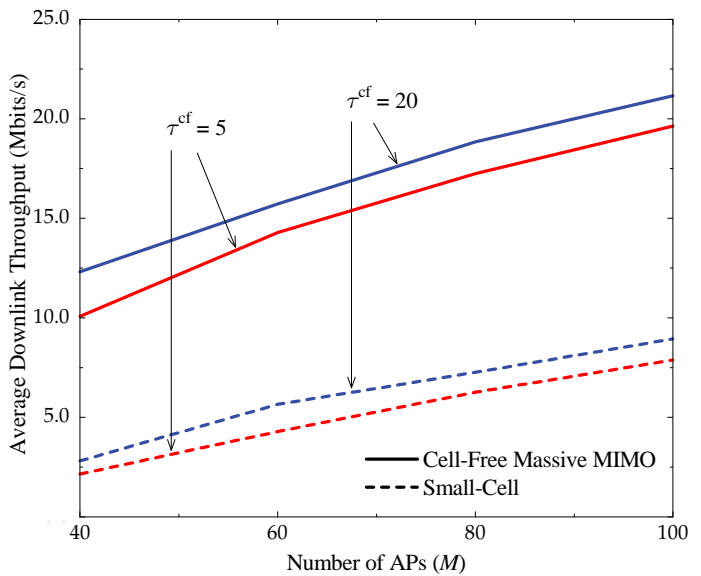}}
\caption{The average downlink net throughput versus the number
of APs for different $\tau^\cf$. Here,
$K=20$.\label{fig:cellfree_smallcell_dl_M}}
\end{figure}

In addition, we study how the $M$ APs assign powers to a given
user in the downlink of Cell-Free Massive MIMO. From
\eqref{eq:xm}, the average transmit power expended by  the $m$th
AP on the $k$th user is $\Pd^\cf\eta_{mk}\gamma_{mk}$. Then
\begin{align}
p(m,k) \triangleq \frac{\eta_{mk}\gamma_{mk}}{\sum_{m'=1}^M
\eta_{m'k}\gamma_{m'k}}
\end{align}
is the ratio between the power spent by the $m$th AP on the $k$th
user and the total power collectively spent by all APs on the
$k$th user. Figure~\ref{fig:8} shows the cumulative distribution
of the effective number of APs serving each user, for
$\tau^{\mathrm{cf}}=5$ and 20, and uncorrelated shadow fading.
The effective number of APs serving each user is defined as the
minimum number of APs that contribute at least 95\% of the power
allocated to a given user. This plot was generated as follows: 1)
$200$ random realizations of the AP/user locations and shadow
fading profiles were generated, each with $M=100$ APs and $K=40$
users; 2) for each user $k$ in each realization, we found the
minimum number of APs, say $n$, such that the $n$ largest values
of $\{p(m,k)\}$ sum up to at least 95\% ($k$ is arbitrary here,
since all users have the same statistics); 3) a cumulative
distribution was generated over the $200$ realizations. We can see
that, on average, only about $10$--$20$ of the $100$ APs really
participate in serving a given user. The larger $\taucf$, the less
pilot contamination and the more accurate channel
estimates---hence, more AP points can usefully serve each user.

Finally, we investigate the effect of the number of users $K$, number of APs $M$, and
the training duration $\tau^\cf$ on the performance of Cell-Free
Massive MIMO and small-cell systems.
Figure~\ref{fig:cellfree_smallcell_dl_K} shows the average
downlink net throughput versus $K$ for different $\tau^\cf$, at $M=100$ and uncorrelated shadow fading. The
average is taken over the large-scale fading. We can see that when
reducing $K$ or $\tau^\cf$, the effect of pilot contamination
increases, and hence, the performance  decreases. As expected,
Cell-Free Massive MIMO systems outperform small-cell systems.
Cell-Free Massive MIMO benefits from favorable propagation, and
therefore, it suffers less from interference than the small-cell
system does. As a result, for a fixed $\tau^\cf$, the relative performance
gap between Cell-Free Massive MIMO and small-cell systems
increases with $K$. Figure~\ref{fig:cellfree_smallcell_dl_M} shows the average
downlink net throughput versus $M$ for different $\tau^\cf$, at $K=20$. Owing to the array gain (for Cell-Free Massive MIMO systems) and diversity gain (for small-cell systems), the system performances of both Cell-Free Massive MIMO and small-cell systems increase when $M$ increases. Again, for all $M$, Cell-Free Massive MIMO systems are significantly better than small-cell systems. 

Tables~\ref{table:2} and \ref{table:3} summarize the downlink
respectively uplink performances of the Cell-Free Massive MIMO and
small-cell systems, under uncorrelated and correlated shadow
fading.

\begin{table*}[b!]
\hrulefill
    \caption{
      The 95\%-likely per-user net throughput (Mbits/s) of the Cell-Free and small-cell downlink, for $M=100$, $K=40$, and $\tau^\cf=\tauscdl=20$.
    }
    \centerline{
\begin{tabular}{|c||c|c||c|c||c|c|}
\hline
 &\multicolumn{2}{c||}{greedy pilot assignment}&\multicolumn{2}{c||}{greedy pilot assignment}& \multicolumn{2}{c|}{random pilot assignment}\\
  &\multicolumn{2}{c||}{with power control}&\multicolumn{2}{c||}{without power control}& \multicolumn{2}{c|}{with power control}\\
\cline{2-7}
 &uncorrelated & correlated & uncorrelated & correlated  & uncorrelated  & correlated\\
 &shadowing & shadowing & shadowing &  shadowing & shadowing & shadowing\\
\hline\hline
Cell-Free & $14$ & $8.12$ & $5.46$ & $1.58$ & 12.70 & 6.95\\
\hline
Small-cell &$2.08$& $0.83$ & $0.86$ & $0.24$ & 1.37 & 0.54\\
\hline
\end{tabular}}
    \label{table:2}
\end{table*}

\begin{table*}[b!]
\hrulefill
    \caption{
      The 95\%-likely per-user net throughput (Mbits/s) of the Cell-Free and small-cell uplink, for $M=100$, $K=40$, and $\tau^\cf=\tauscul=20$.
    }
    \centerline{
\begin{tabular}{|c||c|c||c|c||c|c|}
\hline
 &\multicolumn{2}{c||}{greedy pilot assignment}&\multicolumn{2}{c||}{greedy pilot assignment}& \multicolumn{2}{c|}{random pilot assignment}\\
  &\multicolumn{2}{c||}{with power control}&\multicolumn{2}{c||}{without power control}& \multicolumn{2}{c|}{with power control}\\
\cline{2-7}
 &uncorrelated & correlated & uncorrelated & correlated  & uncorrelated  & correlated\\
 &shadowing & shadowing & shadowing &  shadowing & shadowing & shadowing\\
\hline\hline
Cell-Free & $6.29$ & $3.55$ & $2.71$ & $0.56$ & 5.54 & 2.26\\
\hline
Small-cell &$2.04$& $0.31$ & $0.76$ & $0.13$ & 1.27 & 0.26\\
\hline
\end{tabular}}
    \label{table:3}
\end{table*}

\section{Conclusion} \label{sec:Conclusion}

We analyzed the performance of Cell-Free Massive MIMO, taking into
account the effects of channel estimation, non-orthogonality of
pilot sequences, and power control. A comparison between Cell-Free
Massive MIMO systems and small-cell systems was also performed,
under uncorrelated and correlated shadow fading.

The results show that Cell-Free Massive MIMO systems can
significantly outperform small-cell systems in terms of
throughput. In particular, Cell-Free systems are much more robust
to shadow fading correlation than small-cell systems. The
95\%-likely per-user throughputs of Cell-Free Massive MIMO with
shadowing correlation are an order of magnitude higher than those
of the small-cell systems.  In terms of implementation complexity,
however, small-cell systems require much less backhaul than
Cell-Free Massive MIMO.

\appendix

\subsection{Proof of Theorem~\ref{theorem rate}} \label{sec app theo1}
To derive the closed-form expression for the achievable rate given
in \eqref{eq:rateexpr1}, we need to compute ${\tt DS}_k$,
$\EX{|{\tt BU}_k|^2}$, and $ \EX{|{\tt UI}_{kk'}|^2}$.
\subsubsection{Compute ${\tt DS}_k$}

Let $\varepsilon_{mk}\triangleq g_{mk} -\hat{g}_{mk}$ be the
channel estimation error. Owing to the properties of MMSE
estimation,  $\varepsilon_{mk}$ and $\hat{g}_{mk}$ are
independent. Thus, we have
\begin{align}\label{eq:proof rate1}
    {\tt DS}_k &=  \sqrt{\Pd^{\mathrm{cf}}}\EX{\sum_{m=1}^M \eta_{mk}^{1/2}(\hat{g}_{mk} + \varepsilon_{mk})\hat{g}_{mk}^\ast}\nonumber\\
    &= \sqrt{\Pd^{\mathrm{cf}}}\sum_{m=1}^M \eta_{mk}^{1/2}\gamma_{mk}.
\end{align}

\subsubsection{Compute $\EX{|{\tt BU}_k|^2}$}
Since the variance of a sum of independent RVs is equal to the sum
of the variances, we have
\begin{align}\label{eq:proof rate2}
&\EX{|{\tt BU}_k|^2}
    =
    {\Pd^{\mathrm{cf}}}\sum_{m=1}^M  \eta_{mk}\EX{\left|g_{mk}\hat{g}_{mk}^\ast-
  \!\EX{g_{mk}\hat{g}_{mk}^\ast}\right|^2}\nonumber\\
  &=
  {\Pd^{\mathrm{cf}}}\sum_{m=1}^M
  \eta_{mk}\left(\EX{\left|g_{mk}\hat{g}_{mk}^\ast
  \right|^2} - |\EX{g_{mk}\hat{g}_{mk}^\ast}|^2\right)\nonumber\\
  &=
  {\Pd^{\mathrm{cf}}}\sum_{m=1}^M
  \eta_{mk}\left(\EX{\left|\varepsilon_{mk}\hat{g}_{mk}^\ast + |\hat{g}_{mk}|^2
  \right|^2} - \gamma_{mk}^2\right)\nonumber\\
  &\mathop  = \limits^{(a)}
  {\Pd^{\mathrm{cf}}}\sum_{m=1}^M
  \eta_{mk}\left(\EX{\left|\varepsilon_{mk}\hat{g}_{mk}^\ast
  \right|^2} + \EX{|\hat{g}_{mk}|^4}- \gamma_{mk}^2\right)
  \nonumber\\
  &\mathop  = \limits^{(b)}
  {\Pd^{\mathrm{cf}}}\sum_{m=1}^M
  \eta_{mk}\left(\gamma_{mk}(\beta_{mk}-\gamma_{mk}) + 2\gamma_{mk}^2-
  \gamma_{mk}^2\right)\nonumber\\
  &= {\Pd^{\mathrm{cf}}}\sum_{m=1}^M
  \eta_{mk} \gamma_{mk}\beta_{mk},
\end{align}
where $(a)$ follows that fact that $\varepsilon_{mk}$ has zero
mean and is independent of $\hat{g}_{mk}$, while $(b)$ follows
from the facts that $\EX{|\hat{g}_{mk}|^4} = 2\gamma_{mk}^2$ and
$\EX{|\varepsilon_{mk}|^2}= \beta_{mk}-\gamma_{mk}$.

\subsubsection{Compute $\EX{|{\tt UI}_{kk'}|^2}$}
From \eqref{eq:MMSE est1} and \eqref{eq:rat2c}, we have
\begin{align}\label{eq:proof rate3}
\EX{|{\tt
UI}_{k'}|^2}&=\Pd^{\mathrm{cf}}\E\left\{\rule{0cm}{0.7cm}\left|\sum_{m=1}^M\eta_{mk'}^{1/2}c_{mk'}g_{mk} \right.\right.\nonumber\\&\hspace{-0.5cm}\left.\left.\times\left(\sqrt{\taucf
\Pp^{\mathrm{cf}}}\sum_{i=1}^K g_{mi} \pmb{\varphi}_{k'}^H
\pmb{\varphi}_{i} + \tilde{w}_{mk'}
\right)^\ast\right|^2\right\},
\end{align}
where $\tilde{w}_{mk'}\triangleq \pmb{\varphi}_{k'}^H\B{w}_{\p,m}$
$\sim \CG{0}{1}$. Since $\tilde{w}_{mk'}$ is independent of
$g_{mi}$, $\forall i, k'$, we have
\begin{align}\label{eq:proof rate4}
&\EX{|{\tt UI}_{kk'}|^2} =
\Pd^{\mathrm{cf}}\EX{\left|\sum_{m=1}^M\eta_{mk'}^{1/2}c_{mk'}g_{mk}
\tilde{w}_{mk'}^\ast\right|^2}\nonumber\\
 &+\taucf\Pp^{\mathrm{cf}}\Pd^{\mathrm{cf}}\EX{\left|\sum_{m=1}^M
\eta_{mk'}^{1/2}c_{mk'}g_{mk} \left(\sum_{i=1}^K g_{mi}
\pmb{\varphi}_{k'}^H \pmb{\varphi}_{i} \right)^\ast\right|^2}.
\end{align}
Using the fact that if $X$ and $Y$ are two independent RVs and
$\EX{X}=0$, then $\EX{|X+Y|^2}=\EX{|X|^2}+\EX{|Y|^2}$,
\eqref{eq:proof rate4} can be rewritten as follows
\begin{align}\label{eq:proof rate5}
&\EX{|{\tt UI}_{kk'}|^2} = \Pd^{\mathrm{cf}}\sum_{m=1}^M\eta_{mk'}c_{mk'}^2\beta_{mk}
+\taucf\Pp^{\mathrm{cf}}\Pd^{\mathrm{cf}}(\mathcal{T}_1+\mathcal{T}_2),
\end{align}
where
\begin{align}\label{eq:proof rate6a}
\mathcal{T}_1
    &\triangleq
        \EX{\left|\sum_{m=1}^M
\eta_{mk'}^{1/2}c_{mk'}|g_{mk}|^2 \pmb{\varphi}_{k'}^H
\pmb{\varphi}_{k} \right|^2},\\\label{eq:proof rate6b}
\mathcal{T}_2
    &\triangleq
        \EX{\left|\sum_{m=1}^M
\eta_{mk'}^{1/2}c_{mk'}g_{mk} \left(\sum_{i\neq k}^K g_{mi}
\pmb{\varphi}_{k'}^H \pmb{\varphi}_{i} \right)^\ast\right|^2}.
\end{align}
We first compute $\mathcal{T}_1$. We have
\begin{align}\label{eq:proof rate7}
\mathcal{T}_1
    &=
        \left|\pmb{\varphi}_{k'}^H
\pmb{\varphi}_{k}\right|^2\EX{\sum_{m=1}^M\!\sum_{n=1}^M
\eta_{mk'}^{1/2}\eta_{nk'}^{1/2}c_{mk'}c_{nk'}|g_{mk}|^2
|g_{nk}|^2}\nonumber\\
&= \left|\pmb{\varphi}_{k'}^H
\pmb{\varphi}_{k}\right|^2\EX{\sum_{m=1}^M
\eta_{mk'}c_{mk'}^2|g_{mk}|^4}\nonumber\\&+\left|\pmb{\varphi}_{k'}^H
\pmb{\varphi}_{k}\right|^2\EX{\sum_{m=1}^M\sum_{n\neq m}^M
\eta_{mk'}^{1/2}\eta_{nk'}^{1/2}c_{mk'}c_{nk'}|g_{mk}|^2
|g_{nk}|^2}\nonumber\\
&= 2\left|\pmb{\varphi}_{k'}^H
\pmb{\varphi}_{k}\right|^2\sum_{m=1}^M \eta_{mk'}c_{mk'}^2
\beta_{mk}^2\nonumber\\&+\left|\pmb{\varphi}_{k'}^H
\pmb{\varphi}_{k}\right|^2\sum_{m=1}^M\sum_{n\neq m}^M
\eta_{mk'}^{1/2}\eta_{nk'}^{1/2}c_{mk'}c_{nk'}\beta_{mk}
\beta_{nk}.
\end{align}
Similarly, we have
\begin{align}\label{eq:proof rate8}
\mathcal{T}_2
    &=
    \sum_{m=1}^M\sum_{i\neq k}^K
        \eta_{mk'}c_{mk'}^2\beta_{mk} \beta_{mi} \left|\pmb{\varphi}_{k'}^H \pmb{\varphi}_{i}\right|^2.
\end{align}
Substitution of \eqref{eq:proof rate7} and \eqref{eq:proof rate8}
into \eqref{eq:proof rate5} yields
\begin{align}\label{eq:proof rate9}
\EX{|{\tt UI}_{kk'}|^2}
    &=
    \Pd^{\mathrm{cf}}\left|\pmb{\varphi}_{k'}^H \pmb{\varphi}_{k}\!\right|^2
    \left(\sum_{m=1}^M\eta_{mk'}^{1/2}\gamma_{mk'}\frac{\beta_{mk}}{\beta_{mk'}} \right)^2 \nonumber\\&+ \Pd^{\mathrm{cf}} \sum_{m=1}^M\eta_{mk'}\gamma_{mk'} \beta_{mk}.
\end{align}

Plugging \eqref{eq:proof rate1}, \eqref{eq:proof rate2}, and
\eqref{eq:proof rate9} into \eqref{eq:rateexpr1}, we obtain
\eqref{eq:Theo_rateexpr1}.

\subsection{Proof of Proposition~\ref{prop1} } \label{sec app propPC}
Denote by $\mathcal{S}\triangleq \{\varsigma_{mk}, \varrho_{k'k},
\vartheta_{m}\}$ the set of variables, and $f(\mathcal{S})$ the
objective function of \eqref{eq opt 2}:
\begin{align}\label{eq:proofprop1}
f(\mathcal{S}) \triangleq  \min_{k=1, \cdots, K}
    \frac{ \left(\sum_{m=1}^M \gamma_{mk}\varsigma_{mk} \right)^2 }{ \sum\limits_{k'\neq k}^K\!\!|\pmb{\varphi}_{k'}^H\pmb{\varphi}_{k}|^2 \varrho_{k'k}^2+\!\! \sum\limits_{m=1}^M\!\!\beta_{mk} \vartheta_{m}^2 +\frac{1}{\Pd^{\mathrm{cf}}} }.
\end{align}

For any $t\in \mathbb{R}_{+}$, the upper-level set of
$f(\mathcal{S})$ that belongs to $\mathcal{S}$ is
\begin{align}\label{eq:proofprop2}
U(f,t)
    &=
        \left\{\mathcal{S}: f(\mathcal{S}) \geq t \right\}\nonumber\\
    &=
        \left\{\!\mathcal{S}\!: \frac{ \left(\sum_{m=1}^M \gamma_{mk}\varsigma_{mk} \right)^2 }{ \sum\limits_{k'\neq k}^K\!\!|\pmb{\varphi}_{k'}^H\pmb{\varphi}_{k}|^2 \varrho_{k'k}^2+\!\! \sum\limits_{m=1}^M\!\!\beta_{mk} \vartheta_{m}^2 \!+\!\frac{1}{\Pd^{\mathrm{cf}}} } \!\geq t, ~ \!\forall k \!\right\}\nonumber\\
    &=
        \left\{\mathcal{S}: \left\|\B{v}_k\right\| \leq \frac{1}{\sqrt{t}}\sum_{m=1}^M \gamma_{mk}\varsigma_{mk}, ~ \forall k
        \right\},
\end{align}
where $\B{v}_k\triangleq \left[\B{v}_{k1}^T\B{I}_{-k} \quad
\B{v}_{k2}^T \quad \frac{1}{\sqrt{\Pd^{\mathrm{cf}}}}\right]^T$,
and where $\B{v}_{k1}\triangleq \left[
\pmb{\varphi}_{1}^H\pmb{\varphi}_{k}\varrho_{1k} ~ ... ~
\pmb{\varphi}_{K}^H\pmb{\varphi}_{K}\varrho_{Kk}\right]^T$,
$\B{I}_{-k}$ is a $K\times (K-1)$ matrix obtained from the
$K\times K$ identity matrix with the $k$th column is removed, and
$\B{v}_{k2}\triangleq \left[\sqrt{\beta_{1k}}\vartheta_{1} ~ ... ~
\sqrt{\beta_{Mk}}\vartheta_{M}\right]^T$.

Since the upper-level set $U(f,t)$ can be represented as a SOC, it
is a convex set. Thus, $f(\mathcal{S})$ is quasi-concave.
Furthermore, the optimization problem \eqref{eq opt 2} is a
quasi-concave  optimization problem since the constraint set in
\eqref{eq opt 2} is also convex.


\bibliographystyle{IEEEtran}

\begin{IEEEbiography}
{Hien Quoc Ngo}  received the B.S. degree in electrical engineering from Ho Chi Minh City University of Technology, Vietnam, in 2007. He then received the M.S. degree in Electronics and Radio Engineering from Kyung Hee University, Korea, in 2010, and the Ph.D. degree in communication systems from Link\"oping University (LiU), Sweden, in 2015. From May to December 2014, he visited Bell Laboratories, Murray Hill, New Jersey, USA.

Hien Quoc Ngo is currently a postdoctoral researcher of the Division for Communication Systems in the Department of Electrical Engineering (ISY) at Link\"oping University, Sweden. He is also a Visiting Research Fellow at the School of Electronics, Electrical Engineering and Computer Science, Queen's University Belfast, U.K. His current research interests include massive (large-scale) MIMO systems and cooperative communications.

Dr. Hien Quoc Ngo received the IEEE ComSoc Stephen O. Rice Prize in Communications Theory in 2015. He also received the IEEE Sweden VT-COM-IT Joint Chapter Best Student Journal Paper Award in 2015. He was an \emph{IEEE Communications Letters} exemplary reviewer for 2014, an \emph{IEEE Transactions on Communications} exemplary reviewer for 2015. He has been a member of Technical Program Committees for several IEEE conferences such as ICC, Globecom, WCNC, VTC, WCSP, ISWCS, ATC, ComManTel.
\end{IEEEbiography}

\begin{IEEEbiography}
  {Alexei Ashikhmin} is a Distinguished Member of Technical Staff in
  the Communications and Statistical Sciences Research Department of Bell
  Labs, Murray Hill, New Jersey. His research interests include
  communications theory, massive MIMO, classical and quantum
  information theory, and error correcting codes.

  In 2014 Dr. Ashikhmin received Thomas Edison Patent Award for
  Patent on Massive MIMO System with Decentralized Antennas.
  In 2004 he received the Stephen O. Rice Prize for the best
  paper of IEEE Transactions on Communications.
  In 2002, 2010, and 2011 he received Bell Laboratories
  President Awards for breakthrough research in communication projects.

  From 2003 to 2006, and 2011 to 2014
  Dr. Ashikhmin served as an Associate Editor for IEEE Transactions on
  Information Theory. 
\end{IEEEbiography}

\begin{IEEEbiography}
{Hong Yang} is a member of technical staff at Nokia Bell Labs' Mathematics of Networks and Communications Research Department in Murray Hill, New Jersey, where he conducts research in communications networks. He has previously worked in both the Systems Engineering Department and the Wireless Design Center at Lucent and Alcatel-Lucent, and for a start-up network technology company. He has co-authored many research papers in wireless communications, applied mathematics, and financial economics, co-invented many U.S. and international patents, and co-authored a book Fundamentals of Massive MIMO (Cambridge, 2016). He received his Ph.D. in applied mathematics from Princeton University, Princeton, New Jersey.
\end{IEEEbiography}

\begin{IEEEbiography}
{Erik G. Larsson} is Professor of Communication Systems
at Link\"oping University (LiU) in Link\"oping, Sweden. He 
previously worked for the Royal Institute of
Technology (KTH) in Stockholm, Sweden, the
University of Florida, USA, the George Washington University, USA,
and Ericsson Research, Sweden.  In 2015 he was
a Visiting Fellow at Princeton University, USA, for four months. He received his Ph.D. degree from Uppsala University,
Sweden, in 2002.  

His main professional interests are within the areas of wireless
communications and signal processing. He has co-authored some 130 journal papers
on these topics, he is co-author of the two Cambridge University Press textbooks \emph{Space-Time
Block Coding for Wireless Communications} (2003) and \emph{Fundamentals of Massive MIMO} (2016).
 He is co-inventor on 16 issued and many pending patents on wireless technology.

He served as Associate Editor for, among others, the \emph{IEEE Transactions on
Communications} (2010-2014) and \emph{IEEE Transactions on Signal Processing} (2006-2010). 
During 2015--2016 he served as   chair of the IEEE Signal Processing Society SPCOM technical committee,
and in 2017 he is the past chair of this committee.
He served chair of the steering committee for the \emph{IEEE Wireless
Communications Letters} in 2014--2015.  He was the General Chair of the Asilomar Conference
on Signals, Systems and Computers in 2015, and Technical Chair in
2012.  He is a member of  the IEEE Signal Processing Society Awards Board during 2017--2019.

He received the \emph{IEEE Signal Processing Magazine} Best Column Award twice, in 2012 and 2014, and  
 the IEEE ComSoc Stephen O. Rice Prize in Communications Theory in 2015.  
  He is a Fellow of the IEEE.

\end{IEEEbiography}

\begin{IEEEbiography}
{Thomas L. Marzetta} was born in Washington, D.C. He received the PhD and SB in Electrical Engineering from Massachusetts Institute of Technology in 1978 and 1972, and the MS in Systems Engineering from University of Pennsylvania in 1973. After careers in petroleum exploration at Schlumberger-Doll Research and defense research at Nichols Research Corporation, he joined Bell Labs in 1995 where he is currently a Bell Labs Fellow. Previously he directed the Communications and Statistical Sciences Department within the former Mathematical Sciences Research Center.

Dr. Marzetta is on the Advisory Board of MAMMOET (Massive MIMO for Efficient Transmission), an EU-sponsored FP7 project, and he was Coordinator of the GreenTouch Consortium's Large Scale Antenna Systems Project. He has received awards including the 2015 IEEE Stephen O. Rice Prize, the 2015 IEEE W. R. G. Baker Award, and the 2013 IEEE Guglielmo Marconi Prize Paper Award. He was elected a Fellow of the IEEE in 2003, and he received an Honorary Doctorate from Link\"oping University in 2015.
\end{IEEEbiography}

\end{document}